\documentclass[aps,prb,twocolumn,superscriptaddress,showpacs]{revtex4-1}
\usepackage{graphicx}
\usepackage{dcolumn}
\usepackage{bm}
\usepackage{color,soul}
\usepackage{amsmath}
\usepackage{braket}
\usepackage{csquotes}
\usepackage{txfonts}
\usepackage{amssymb}

\begin{document}

\preprint{APS/123-QED}
\title{The nexus between negative charge-transfer and reduced on-site Coulomb energy in a correlated topological metal CoTe$_2$}

\author{A. R. Shelke}
\affiliation{National Synchrotron Radiation Research Center, Hsinchu 300092, Taiwan}
\author{C.-W. Chuang}
\affiliation{Experimentelle Physik VII, Universit\"{a}t W\"{u}rzburg, Am Hubland, D-97074 W\"{u}rzburg, Germany}
\author{S. Hamamoto}
\affiliation{RIKEN SPring-8 Center, Hyogo 679-5148, Japan}
\author{M. Oura}
\affiliation{RIKEN SPring-8 Center, Hyogo 679-5148, Japan}
\author{M. Yoshimura}
\affiliation{National Synchrotron Radiation Research Center, Hsinchu 300092, Taiwan}
\author{N. Hiraoka}
\affiliation{National Synchrotron Radiation Research Center, Hsinchu 300092, Taiwan}
\author{C.-N. Kuo}
\affiliation{Program on Key Materials, Academy of Innovative Semiconductor and Sustainable Manufacturing (AISSM), National Cheng Kung University, Tainan 70101, Taiwan}
\affiliation{Department of Physics, National Cheng Kung University, Tainan 70101, Taiwan}
\affiliation{Taiwan Consortium of Emergent Crystalline Materials, National Science and Technology Council, Taipei, Taiwan}
\author{C.-S. Lue}
\affiliation{Program on Key Materials, Academy of Innovative Semiconductor and Sustainable Manufacturing (AISSM), National Cheng Kung University, Tainan 70101, Taiwan}
\affiliation{Department of Physics, National Cheng Kung University, Tainan 70101, Taiwan}
\affiliation{Taiwan Consortium of Emergent Crystalline Materials, National Science and Technology Council, Taipei, Taiwan}
\author{A. Fujimori}
\affiliation{Department of Physics, The University of Tokyo, 7-3-1 Hongo, Bunkyo-ku, Tokyo 113-0033, Japan}
\affiliation{National Synchrotron Radiation Research Center, Hsinchu 300092, Taiwan}
\affiliation{Center for Quantum Technology, and Department of Physics, National Tsing Hua University, Hsinchu 300044, Taiwan}
\author{A. Chainani}
\affiliation{National Synchrotron Radiation Research Center, Hsinchu 300092, Taiwan}

\date{\today}
\begin{abstract}
The layered $3d$ transition metal dichalcogenide (TMD)  CoTe$_2$ is a topological Dirac Type-II metal. However, the Co $3d$-bands in CoTe$_2$ do not exhibit the expected correlation-induced band narrowing seen in CoO. We address this conundrum by studying the electronic structure of CoTe$_2$ using hard x-ray photoemission spectroscopy (HAXPES), x-ray absorption spectroscopy (XAS) and  Resonant-PES. We quantify the on-site Coulomb energy $U_{dd}$ via single-particle partial density of states and the two-hole correlation satellite using valence band Resonant-PES), and obtain $U_{dd}$ = 3.0 eV for CoTe$_2$. Charge-transfer (CT) cluster model simulations of the experimental core-level Co $2p$ PES and $L$-edge XAS spectra of CoTe\textsubscript{2} and CoO validate their contrasting electronic parameters:$U_{dd}$ and CT energy $\Delta$ are (3.0 eV, -2.0 eV) for CoTe\textsubscript{2}, and (5.0 eV, 4.0 eV) for CoO, respectively. The $d$-$p$ hybridization strength $T_{eg}$ for CoTe$_2$$<$CoO, and indicates that the reduced $U_{dd}$ in CoTe\textsubscript{2} is not due to $T_{eg}$. The increase in $d^n$-count$\sim$1 by CT from ligand to Co site in CoTe$_2$ is due to a negative-$\Delta$ and reduced $U_{dd}$. Yet, only because $U_{dd}$$>$$\big|\Delta\big|$, CoTe$_{2}$ becomes a topological metal  with \textit{p}$\rightarrow$\textit{p} type lowest energy excitations. The study reveals the nexus between negative-$\Delta$ and reduced $U_{dd}$ required for setting up the electronic structure framework for achieving  topological behavior via band inversion in the correlated metal CoTe$_2$.
\end{abstract}

\maketitle

\section{\label{sec:levelI}Introduction}
The $3d$ transition metal compounds (TMCs) show the largest variety of physical and chemical properties as $3d$ electrons manifest localized/delocalized as well as magnetic/non-magnetic character\cite{Imada1998}. Early studies\cite{Adler,Mattheiss} considered TMC properties to originate purely from $d$-electrons, and ligand $p$-states were neglected assuming they were fully-filled bands lying far below the Fermi level($E_F$). The work of Fujimori and Minami,\cite{FujimoriNiO} and that of Zaanen, Sawatzky and Allen(ZSA)\cite{ZSA} led to a paradigm shift by considering ligand $p$-electron states on an equal footing with $d$-electron states for describing the ground state character as well as excitations in TMCs. TMCs are now mainly characterized by the metal $d$ on-site Coulomb energy $U_{dd}$ and charge-transfer energy $\Delta$ between $d$ and $p$ states, both scaled by the $d$-$p$ hybridization strength $T$. 

The ZSA phase diagram\cite{ZSA} classified TMCs into 4 categories: (i) Mott-Hubbard(MH) insulators with $U_{dd}<\Delta$ and a $d$-$d$ gap between the lower(occupied) and upper(unoccupied) Hubbard bands (LHB and UHB, respectively), (ii) MH $d$-band metals, arising from LHB overlapping UHB, (iii) CT insulators with $U_{dd}>\Delta$ exhibiting a $p$-$d$ gap between the occupied ligand band and UHB, and (iv) a $p$-type metal phase derived from a negative-$\Delta$. Evidence of negative-$\Delta$ was first reported in insulating NaCuO$_2$ with a $p-p$ gap\cite{Mizokawa},
and is now known in several TMCs like NdNiO$_3$\cite{Bisogni}, AuTe$_2$\cite{Streltsov}, TMS$_2$\cite{ana2024}, etc.

Extensive experimental and theoretical studies have shown that variation in electronic and magnetic  properties of $3d$ TMCs originate from a large systematic variation in $U_{dd}$ ($\sim$2-8 eV for Ti-Cu\cite{Imada1998,Mattheiss,FujimoriNiO,ZSA,Zaanen,Anisimov,Casula}) and ligand electronegativity which determines $\Delta$ in solids\cite{Pavarini}. Here,  as a measure of $U_{dd}$, we include empirical values $U_{E}$ (obtained from PES and inverse-PES experimental results\cite{Zaanen}), $U$ as used in density functional theory(DFT) calculations\cite{Anisimov}, and $U_0$, the zero-frequency screened Coulomb interaction\cite{Casula}. In contrast, the $4d$ and $5d$  series of TMCs show much smaller values of $U_{dd}$ ($\sim$1-2 eV), and seemed less susceptible to cause property variations. In fact, until $\sim$2005, only late TM oxides were considered strongly correlated, while early $3d$ and the entire $4d$, $5d$ series of TMCs were often said to be weakly correlated\cite{RSSingh}. 

However, studies on early-$3d$ and $4d$, $5d$ TMCs have now shown that correlation effects can lead to Mott gaps in VTe$_2$\cite{Zhao2023}, Ba$_2$NaOsO$_6$\cite{Erickson}, Sr$_2$IrO$_4$ \cite{BJKim}, Na$_2$IrO$_3$ \cite{Singh}, etc. 
Further, a Mott-insulator phase\cite{Cao3} and superconductivity\cite{Cao2} were discovered in twisted bilayer graphene with a very small effective Coulomb energy $U_{eff}$($\sim$30 meV), but the narrow bandwidth ($W$$\sim$20 meV) of the relevant Moir\'{e} band clarified its Mott-insulator behavior\cite{Cao3}. The above discussions and examples indicate the importance of quantifying and comparing $U_{dd}$,  $\Delta$ and the $d$-$p$ hybridization strength $T$ or  bandwidth $W$ for describing correlation effects. A recent angle-resolved PES (ARPES) study of the late-$3d$ TMD 1$T$-CoTe$_2$ (hereafter, CoTe$_2$) has shown that it is a topological Dirac Type-II metal with bulk spin-orbit split Te $5p$ bands crossing $E_F$\cite{Chakraborty2023}. Furthermore, CoTe$_2$  
shows weak Pauli paramagnetic-type susceptibility, and linear magnetoresistance upto 9 T applied field\cite{WangCM}.
Surprisingly, CoTe$_2$ does not show evidence for the expected strong correlation effects like the very narrow $d$-band dispersions in CoO, which showed only 25\% widths of the band dispersions obtained from DFT calculations\cite{ZXShen}. 
 Thus, it is important to study CoTe$_2$ and quantify the electronic parameters $U_{dd}$, $\Delta$  and $T$ in order to characterize its electronic structure.

In particular, band structure calculations of CoTe$_2$ showed 
band inversion between Te $p_x$+$p_y$ and Te $p_z$ orbitals just below and above $E_F$, leading to a bulk character Type-II Dirac point lying $\sim$0.9 eV above $E_F$\cite{Chakraborty2023}. Further, in combination with ARPES studies, it was shown that the Dirac points in surface states lie $\sim$0.5 eV below $E_F$. However, the bandwidths obtained from DFT-GGA calculations without $U$ were reduced by 70\% to match experimental data, suggesting weak correlations\cite{Chakraborty2023}. Most importantly, the role of $d$-bands and electronic parameters of CoTe$_2$ and specifically, the relation of $\Delta$ and $U_{dd}$  with topological states and properties have not been addressed to date. 

In this work, we use bulk-sensitive HAXPES, XAS and  Resonant-PES to study the electronic structure of single crystal CoTe$_2$.
The synthesis, structural characterization, and details of spectroscopy are described in the Methods.  HAXPES core level and XAS measurements of CoTe$_2$ are then described to identify valency, charge-transfer satellites and plasmon features. We then use resonant PES to quantify $U_{dd}$ in CoTe$_2$ via measurements of the single-particle 3\textit{d} partial density of states (PDOS) and the two-hole correlation satellite using the Cini-Sawatzky method\cite{cini1976,cini1977,sawatzky1977,cini1978}. While ARPES studies corroborating the high-quality of CoTe$_2$ samples (from the same batches as present samples) have already been reported\cite{Chakraborty2023}, soft x-ray valence band PES measurements have not been reported to date. We carry out soft x-ray angle-integrated off- and on-resonant valence band PES ($h\nu$= 700-900 eV, 1.5 keV), as well as bulk-sensitive valence band HAXPES  ($h\nu$= 6.5 keV) to determine the partial density of states(PDOS). Cluster model calculations provide the electronic parameters which indicate CoTe$_2$ is a moderately correlated negative-$\Delta$ material.

\section{\label{sec:levelI}Methods}

\subsection{Synthesis and structure characterization}

The single crystals of CoTe$_2$ were prepared by the chemical vapor transport method, using iodine as the transport agent\cite{Prodan}. High-purity Co (99.95\%) and Te (99.999\%) powders were mixed with a small amount of iodine (40 mg), sealed in an evacuated quartz tube, and then heated for 15 days in a two-zone furnace with a source zone temperature of 825$^\circ$C and a growth zone temperature of 750$^\circ$C. Finally, the quartz tube was quenched into an ice-water bath from the growth temperature of 825$^\circ$C. The obtained single crystals are hexagonal in shape with typical dimensions of 2 mm$\times$2 mm$\times$0.1 mm. 
The crystal structure was characterized using powder X-ray diffraction (XRD) (Bruker D2 phaser diffractometer) with Cu-K$\alpha$ radiation. The single crystal quality was confirmed and crystallization directions were identified by the Laue diffraction method (Photonic Science). 
The XRD results showed a 1T-CdI$_2$-type trigonal structure (space group of P$\bar{3}$m1 (No. 164)) with the flat-plates corresponding to the (001) plane. The obtained lattice parameters of CoTe$_2$ are $a = b$ = 3.791$\AA$ and $c$ = 5.417$\AA$ ($c/a$ = 1.429). These values are very close to reported values of $a = b$ = 3.804$\AA$ and $c$ = 5.401$\AA$ ($c/a$ = 1.421) for CoTe$_2$\cite{TMdatabase}.
 It is known that CoTe$_2$ exhibits a stable divalent state of Co$^{2+}$ ions and the Te atoms are dimerized (Te$_2^{2-}$)\cite{Jobic}. It was shown that the Te-Te distance along the c-axis (3.446$\AA$ and 3.521$\AA$ for CoTe$_2$ were much smaller than the sum of two Te$^{2-}$ ionic radii [r(Te$^{2-}$) = 2.21$\AA$; ref.\cite{Shannon}] due to dimerization. The reduced Te-Te distance along the c-axis results in a significantly reduced $c/a$ ( $\sim$1.27-1.43 ) for many late TM ditellurides compared to early TM ditellurides such as TiTe$_2$, ZrTe$_2$ and HfTe$_2$ which show a much larger $c/a$ ( $\sim$1.68-1.73 )\cite{Jobic}.
 
 \subsection{Electron spectroscopy}
Hard x-ray photoemission spectroscopy (HAXPES) core level and valence band measurements of CoTe$_2$ were carried out at BL-12XU (Taiwan Beamline), SPring8, Japan using linearly polarized X-ray beam with incident photon energy h$\nu$ = 6.5 keV. Liquid He closed cycle cryostat was used to cool the sample, down to a temperature of T = 25 K. The Fermi level \textit{E}\textsubscript{F} of a Au thin film was also measured at T = 25 K to calibrate the binding energy(BE) scale. The total energy resolution is 270 meV as estimated from the fitting of Au \textit{E}\textsubscript{F}. The single crystal samples were cleaved using a top-post in ultra-high vacuum (UHV) of 5.5 $\times$ 10\textsuperscript{-9} mbar in the preparation chamber and then quickly transferred to the main chamber at 7.0 $\times$ 10\textsuperscript{-10} mbar for the measurements. 
Soft x-ray PES (SXPES) core level and valence band, Co $L$-edge XAS and 2\textit{p}-3\textit{d} resonant-PES measurements were carried out at BL-17SU RIKEN beamline in SPring-8, Japan using a circularly polarized x-ray beam. Co $L$-edge XAS of single crystal CoO was measured as a reference compound to confirm the photon energy scale calibration and for comparison with Co $L$-edge XAS of CoTe$_2$. The XAS measurements were carried out in total electron yield mode. SXPES core levels and valence band measurements were carried out with incident photon energy h$\nu$ = 1.5 keV, except for the Co 2\textit{p} and Te 3\textit{p} core levels which were measured with h$\nu$ = 1.7 keV to avoid overlapping Auger lines in the spectrum at h$\nu$ = 1.5 keV. A Liquid N$_2$ flow-type cryostat was used to cool the sample down to 80 K. The total energy resolution at T = 80 K was set to 220-400 meV for h$\nu$ = 700 eV - 1.5 keV, respectively, as confirmed from fits to the Au \textit{E}\textsubscript{F} measured with h$\nu$ = 700 eV and 1.5 keV. Thus, the resolutions of soft x-ray PES (220-400 meV) are comparable to the HAXPES resolution of 270 meV at h$\nu$ = 6.5 keV. The single crystal samples were cleaved and measured in the main chamber at a UHV of 1.0 $\times$ 10\textsuperscript{-10} mbar. 

\subsection{Cluster-model calculations}
The metal 2\textit{p} PES core level and \textit{L}\textsubscript{3,2}-edge XAS spectra  were calculated based on a charge transfer multiplet cluster model using the QUANTY code\cite{haverkort2012, lu2014, haverkort2014}. The divalent ion of Co$^{2+}$ in CoTe$_2$ and CoO correspond to 3$d^{7}$ electronic configuration. A M\textit{L}\textsubscript{6} cluster with octahedral symmetry (\textit{O}\textsubscript{h}), M = Co atoms and \textit{L} is ligand Te atoms. The initial state consists of a linear combination of $\ket{3\textit{d}\textsuperscript{n}}$, $\ket{3\textit{d}\textsuperscript{n+1}\underline{\textit{L}}\textsuperscript{1}}$ and $\ket{3\textit{d}\textsuperscript{n+2}\underline{\textit{L}}\textsuperscript{2}}$ states, where \underline{\textit{L}} corresponds to hole in ligand states. The electronic parameters of the calculation are: the on-site Coulomb energy \textit{U}\textsubscript{dd} (obtained from the Cini-Sawatzky analysis), the charge transfer energy $\Delta$ is the energy difference between $\ket{3\textit{d}\textsuperscript{n}}$ and $\ket{3\textit{d}\textsuperscript{n+1}\underline{\textit{L}}\textsuperscript{1}}$ states, the Co 3\textit{d}-Te 5\textit{p} hybridization strength (T$_{eg}$ and T$_{t2g}$ = T$_{eg}$/2 ) and the crystal field splitting10\textit{Dq} between \textit{t}\textsubscript{2g} and \textit{e}\textsubscript{g} orbitals. For calculating the $2p$ PES spectrum, the final state is a free electron with a core hole in the $2p$ level and includes the on-site core hole potential $U_{pd}$ on the metal site. The final states for XAS correspond to $2p$-$3d$ dipolar excitations from the metal $L$ edge to the unoccupied $d$ states, and also includes $U_{pd}$. In principle, one needs to use a series of excited configurations of the $d^{n}$ state also, but is computationally difficult\cite{deGroot}. In general, it has been found that reducing the 
the Slater integrals to 80\% of their original Hartree-Fock values, the deviation of calculations compared to experimental results can be minimized by applying this semi-empirical correction to the Slater integrals\cite{deGroot}. 
The calculated spectra are obtained by convoluting the discrete final states by broadening it with a Lorentzian function  for 2\textit{p}\textsubscript{3/2} and 2\textit{p}\textsubscript{1/2} lifetimes, respectively) and a Gaussian function for the experimental spectral width.

\section{Results and Discussion}

\subsection{Core level analysis of CoTe$_2$}

Figure 1 shows a wide spectral range covering the Co 2\textit{p} and Te 3\textit{p} core levels, measured with 
HAXPES ($h\nu$ = 6.5 keV) and SXPES($h\nu$ = 1.7 keV). Based on the binding energy(BE) positions of the high intensity peaks, 
the four main peaks of Co 2\textit{p}\textsubscript{3/2}, Co 2\textit{p}\textsubscript{1/2}, Te 3\textit{p}\textsubscript{3/2} and Te 3\textit{p}\textsubscript{1/2} can be suitably assigned\cite{XPShandbook}. Since the Co 2\textit{p} and Te 3\textit{p} core levels have significantly different photoionization cross section (PICS)\cite{trzhaskovskaya2018} with $h\nu$ = 6.5 keV and $h\nu$ = 1.7 keV, the data are normalized at the Te 3\textit{p}\textsubscript{1/2} main peak to see the relative change in Co 2\textit{p} core levels at different photon energies. As is clear from Fig. 1, the SXPES Co 2\textit{p} spectra show a much higher intensity of the main peaks compared to the HAXPES data. In addition to the main peaks, low intensity satellites are observed at higher BEs to the four main peaks. While the low intensity satellites at higher BEs  of the Te 3\textit{p}\textsubscript{3/2} and Te 3\textit{p}\textsubscript{1/2} look very similar in shape and widths for both HAXPES and SXPES data, the low intensity satellites of Co 2\textit{p}\textsubscript{3/2} and Co 2\textit{p}\textsubscript{1/2} seem to show differences in HAXPES and SXPES data. In particular, the satellite of the Co 2\textit{p}\textsubscript{1/2} peak seems to shows higher intensity than the satellite of the Co 2\textit{p}\textsubscript{3/2} peak, and opposite to the behavior of the Te 3\textit{p}\textsubscript{1/2} satellite which shows slightly  lower intensity compared to the Te 3\textit{p}\textsubscript{3/2} satellite. Further, the Co 2\textit{p}\textsubscript{3/2} and Co 2\textit{p}\textsubscript{1/2} satellites in the HAXPES data seem to show very small intensity compared to the SXPES data.

\begin{figure}
\includegraphics[scale=0.32]{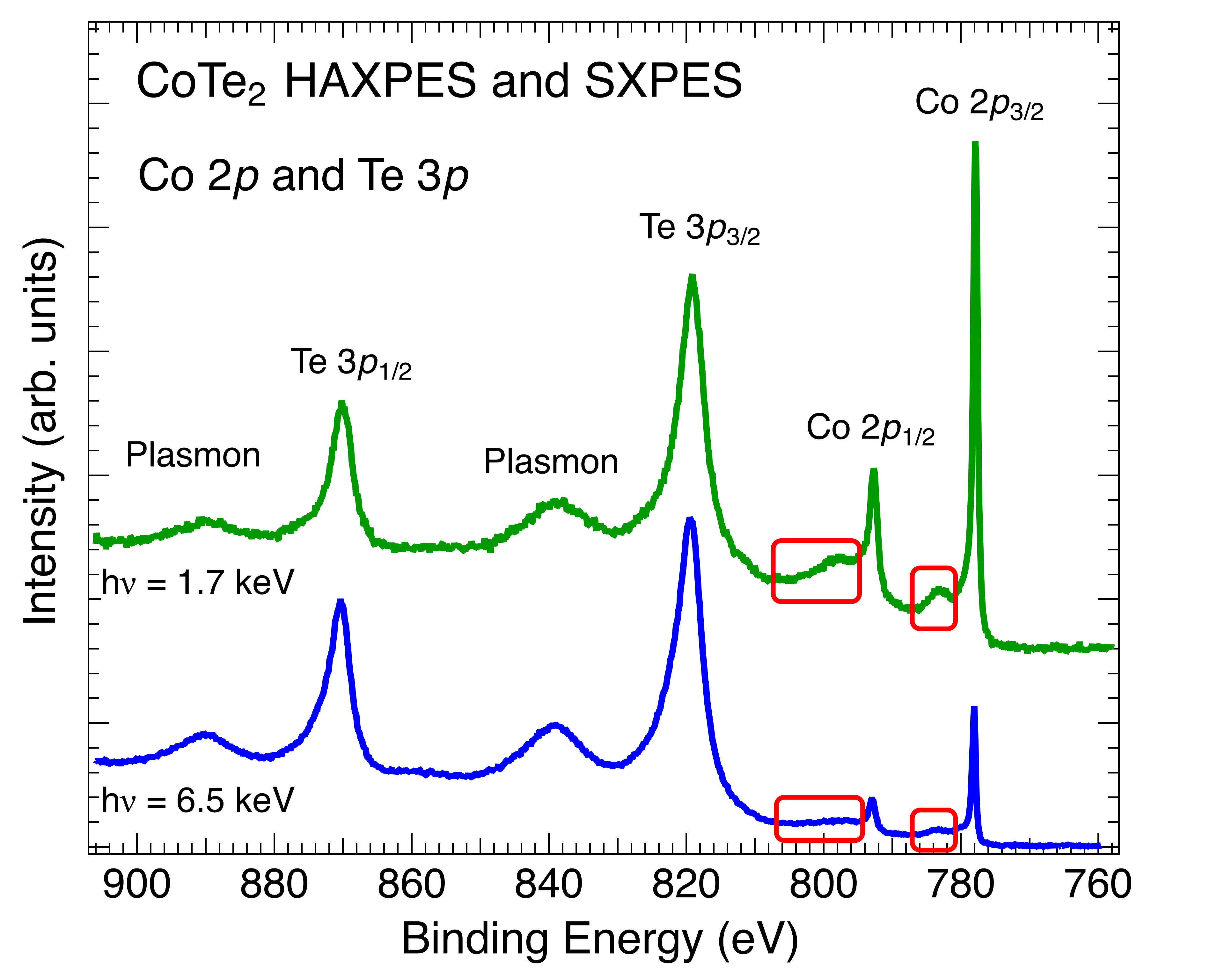}
\caption{Co 2\textit{p} and Te 3\textit{p} core level spectra of CoTe\textsubscript{2} single crystal measured at T =20 K with h$\nu$ = 6.5 keV (HAXPES) and at T =80 K with h$\nu$ = 1.7 keV (SXPES)}
\end{figure}

\begin{figure}
\includegraphics[scale=0.32]{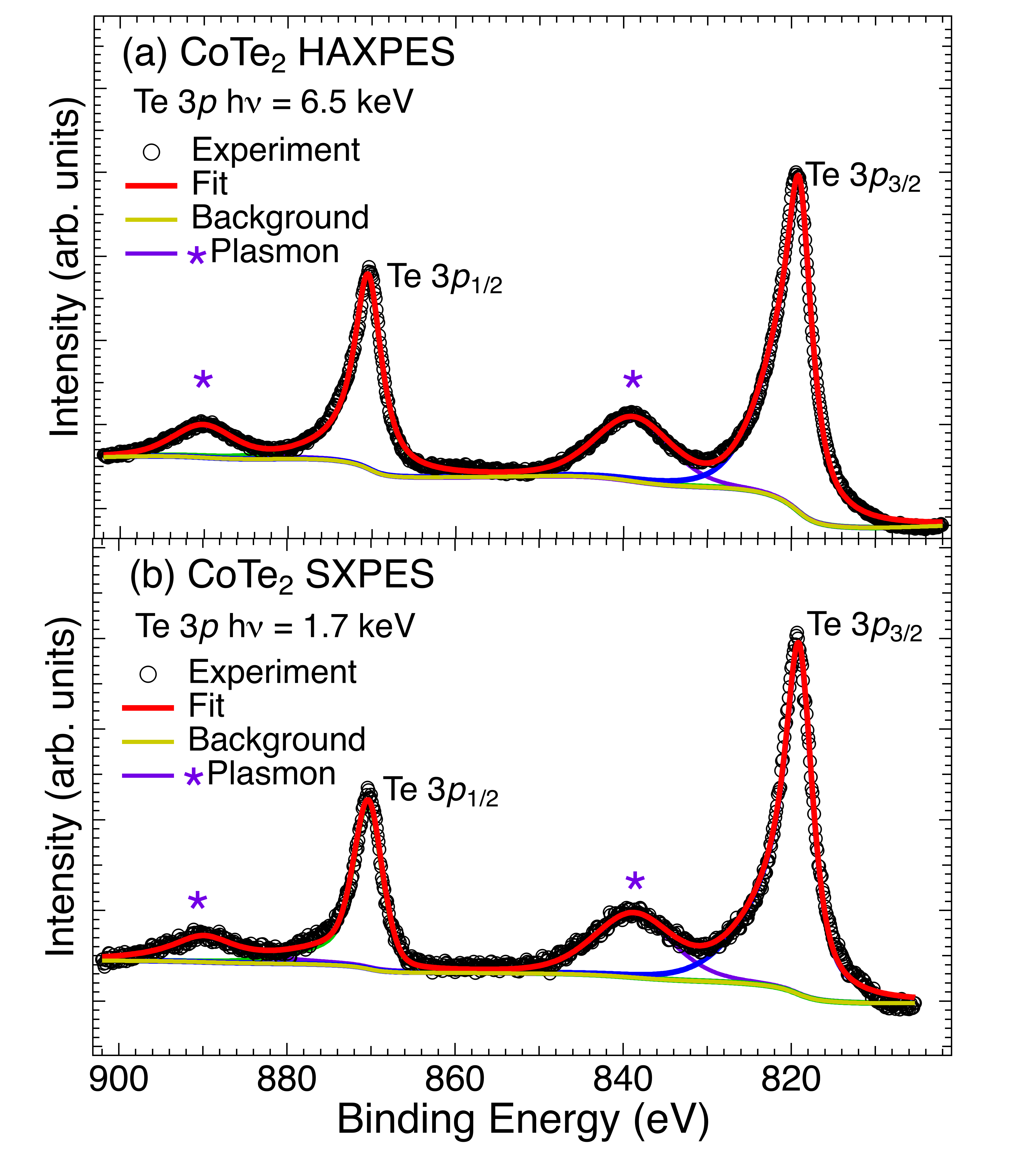}
\caption{Least-squares fitting of Te 3\textit{p} core levels of CoTe\textsubscript{2} single crystal measured using (a) HAXPES and (b) SXPES techniques}
\end{figure}

In order to quantify the peak energy positions and shapes, and to clarify the character and difference between the Co 2\textit{p}\textsubscript{3/2} and 2\textit{p}\textsubscript{1/2 } main peaks and satellites, as well as for Te 3\textit{p}\textsubscript{3/2} and 3\textit{p}\textsubscript{1/2} main peaks and satellites, we first carried out a least-squares fitting of the simpler case of Te 3\textit{p}\textsubscript{3/2} and Te 3\textit{p}\textsubscript{1/2} HAXPES and SXPES core levels. The main peaks could be fitted with single asymmetric Voigt functions typical of metals, while the satellites required a symmetric Gaussian function. The results are shown in Fig. 2(a) and 2(b) with the fits (full lines) overlaid on the experimental spectra (empty circles). 
The peak energy positions and peak full-widths and half maximum (FWHM) are listed in Table I.
The main peak BEs are very similar: 819.06 eV and 870.16 eV in HAXPES and 819.12 eV and 870.24 eV in SXPES,
and the separation in both cases (= 51.1$\pm$0.1eV) is quite close to the known Te 3\textit{p}\textsubscript{3/2} and Te 3\textit{p}\textsubscript{1/2} spin-orbit splitting of 51.0 eV.\cite{XPShandbook} The observed BE values are slightly lower compared to the elemental Te 3\textit{p}\textsubscript{3/2} and Te 3\textit{p}\textsubscript{1/2} values.\cite{XPShandbook} The broad satellites in HAXPES and SXPES data are positioned at 19.8$\pm$0.2 eV higher BE to Te 3\textit{p}\textsubscript{3/2} and Te 3\textit{p}\textsubscript{1/2} main peaks, and suggest a plasmon origin of the satellites.

\begin{table}[t!]
	\begin{center}
	\caption{Fitting parameters of Te 3\textit{p} core levels of CoTe\textsubscript{2} single crystal measured using HAXPES and SXPES}
\begin{ruledtabular}
\begin{tabular}{ccc}
 Fit component  & Binding Energy   & FWHM \\
 HAXPES &          (eV) &                   (eV) \\
\hline
Te 3$p_{3/2}$ & 819.06 & 3.72 \\
Te 3$p_{1/2}$ & 870.16 & 3.89 \\
 Plasmon & 838.86 & 10.50 \\
Plasmon & 890.15 & 8.33 \\
\hline
SXPES \\
Te 3$p_{3/2}$ & 819.12 & 3.83 \\
Te 3$p_{1/2}$ & 870.24 & 3.97 \\
Plasmon & 838.79 & 11.04 \\
Plasmon & 889.96 & 9.50 \\
\end{tabular}
\end{ruledtabular}
\end{center}
\end{table}

\begin{figure}
\includegraphics[scale=0.32]{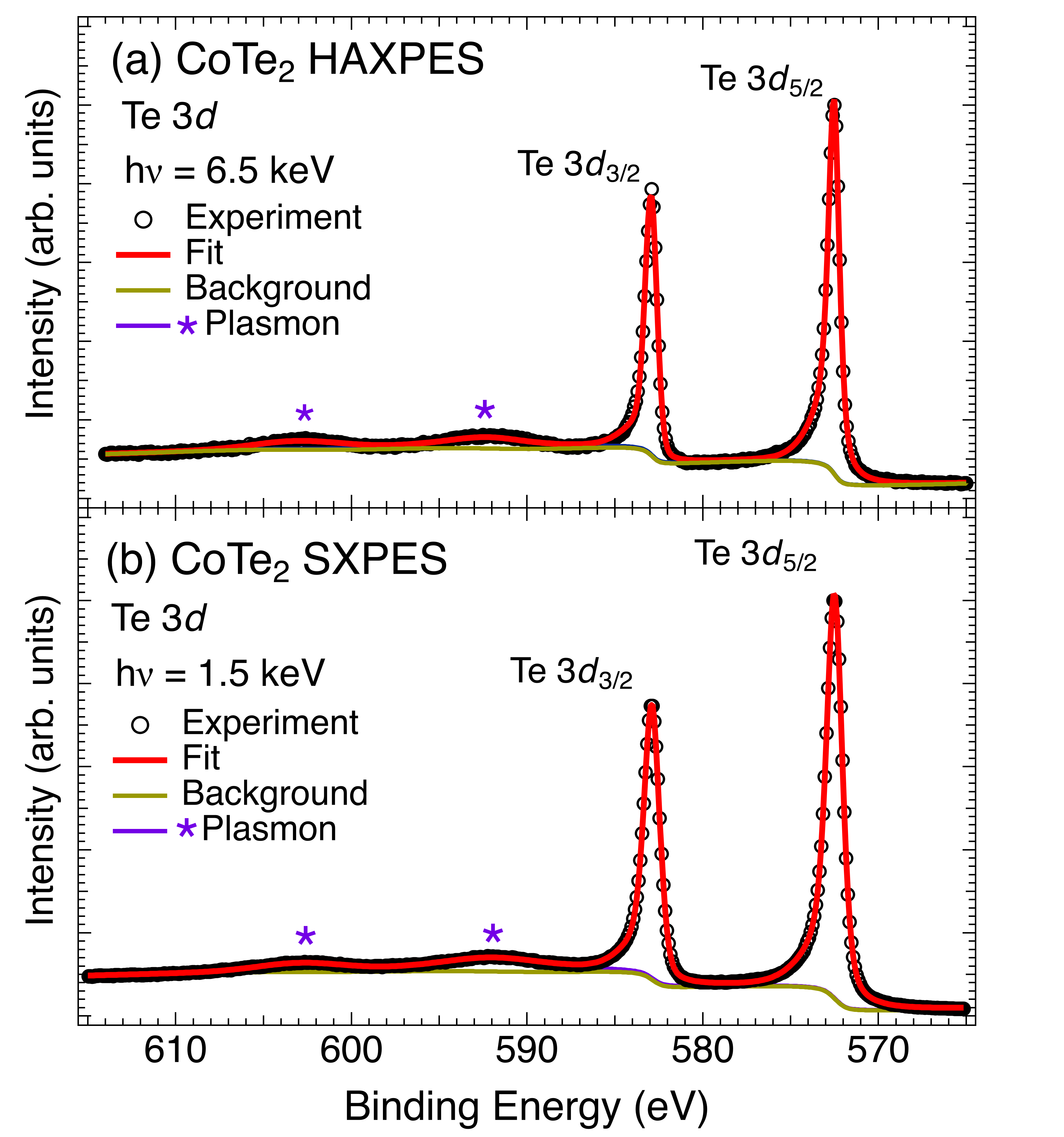}
\caption{Least-squares fitting of Te 3\textit{d} core level CoTe\textsubscript{2} single crystal measured using (a) HAXPES and (b) SXPES techniques }
\end{figure}

\begin{table}[t!]
	\begin{center}
	\caption{Fitting parameters of Te 3\textit{d} core level of CoTe\textsubscript{2} using HAXPES and SXPES}
\begin{ruledtabular}
\begin{tabular}{ccc}
 Fit component  & Binding Energy   & FWHM \\
 HAXPES &          (eV) &                   (eV) \\
\hline
Te 3$d_{5/2}$ & 572.53 & 0.76  \\
Te 3$d_{3/2}$ & 582.94 & 0.80 \\
 Plasmon & 592.20 & 6.50 \\
Plasmon & 602.57 & 6.00 \\
\hline
SXPES \\
Te 3$d_{5/2}$ & 572.50 & 1.03 \\
Te 3$d_{3/2}$ & 582.89 & 1.06 \\
Plasmon & 592.15 & 8.01 \\
Plasmon & 602.52 & 5.90 \\
\end{tabular}
\end{ruledtabular}
\end{center}
\end{table}

\begin{figure}
\includegraphics[scale=0.31]{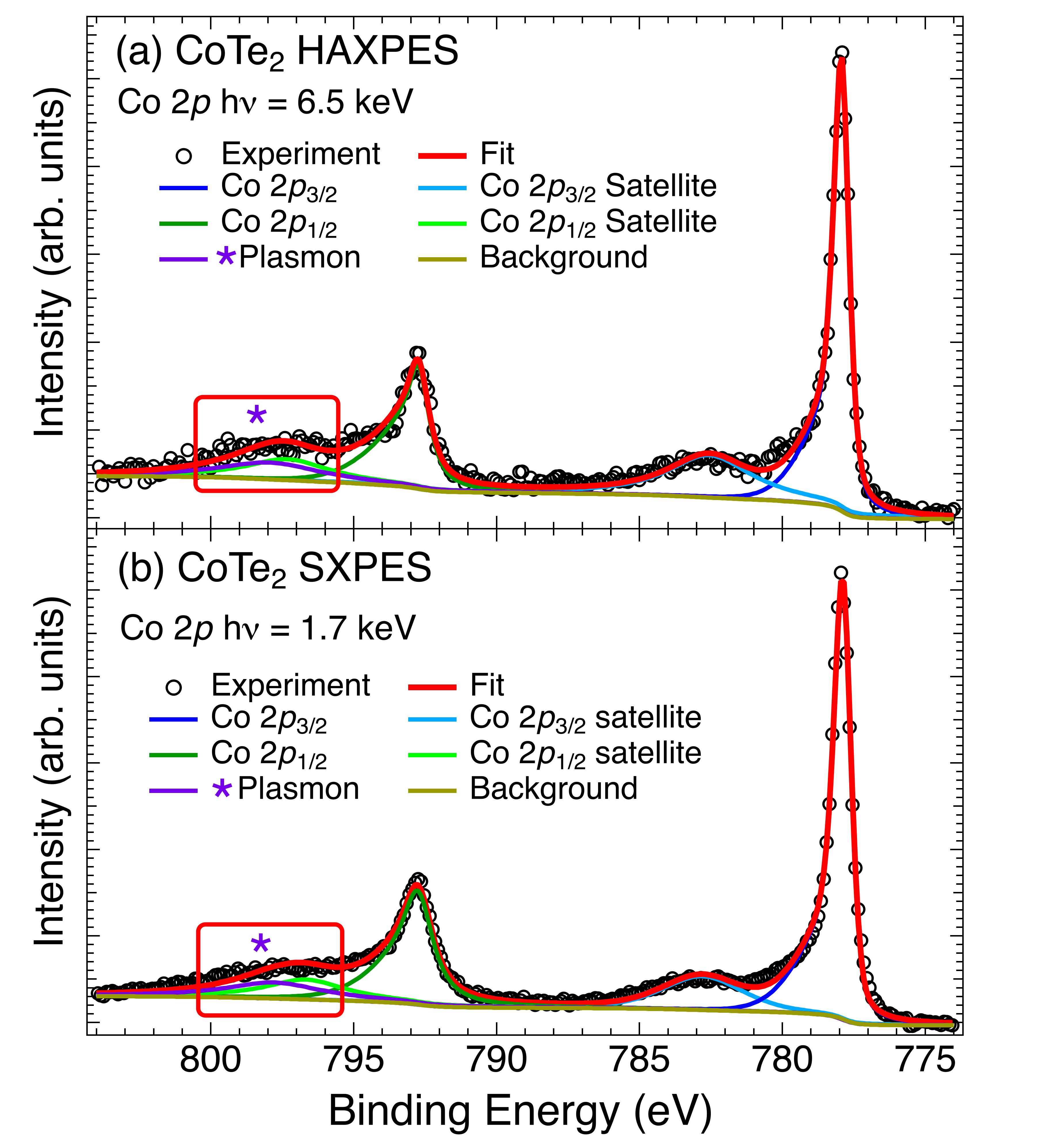}
\caption{Least square fitting of Co 2\textit{p} core levels of CoTe\textsubscript{2} single crystal measured using (a) HAXPES and (b) SXPES techniques to separate out the plasmon of Co 2\textit{p}\textsubscript{3/2} core level lying at a BE close to the Co 2\textit{p}\textsubscript{1/2} core level satellite.}
\end{figure}

\begin{table}[t!]
	\begin{center}
	\caption{Fitting parameters of Co 2\textit{p} core levels of CoTe\textsubscript{2} single crystal measured using HAXPES and SXPES}
\begin{ruledtabular}
\begin{tabular}{ccc}
 Fit component  & Binding Energy   & FWHM \\
 HAXPES &          (eV) &                   (eV) \\
\hline
Co 2$p_{3/2}$ & 778.06 & 0.63  \\
Co 2$p_{1/2}$ & 792.86 & 0.95 \\
Co 2$p_{3/2}$ Satellite &783.01 & 3.33\\
Co 2$p_{1/2}$ Satellite &795.47 & 3.21\\
Co 2$p_{3/2}$ Plasmon & 797.82 & 4.31\\
\hline
SXPES \\
Co 2$p_{3/2}$ & 777.90 & 0.70  \\
Co 2$p_{1/2}$ & 792.70 & 1.14 \\
Co 2$p_{3/2}$ Satellite &783 & 3.36\\
Co 2$p_{1/2}$ Satellite  &795.65 & 3.30\\
Co 2$p_{3/2}$ Plasmon & 797.76 & 4.44 \\
\end{tabular}
\end{ruledtabular}
\end{center}
\end{table}

\begin{figure*}
\centering
    \includegraphics[width=1.0\textwidth]{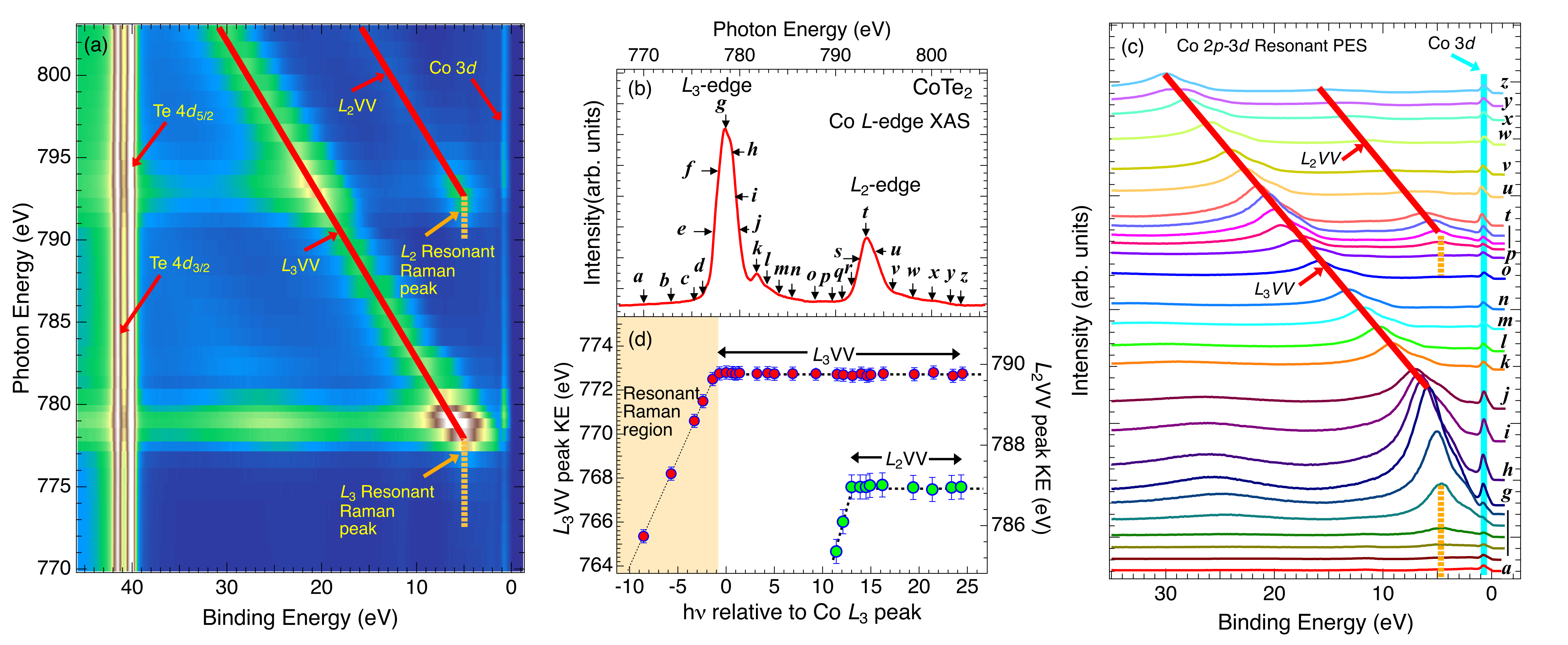}
    \caption{(a). The Co 2\textit{p}-3\textit{d} resonant-PES valence band intensity map of CoTe$_2$ plotted as a function of incident photon energies ($h\nu$ = 770-803 eV) versus binding energy (BE = -1.2 to 45.8 eV). (b) The Co $L_3$- and $L_2$-edge XAS of CoTe$_2$ plotted as a function of $h\nu$(top x-axis). (c) Valence band spectra (BE = -1.2 to 35.0 eV) of CoTe$_2$  measured at select $h\nu$ values (labelled $a$-$z$) across the $L_3$- and $L_2$-edges of Fig. 1(b). (d) The kinetic energy of the Resonant Raman - $L_3VV$ Auger peak plotted as a function of $h\nu$(top axis), and also relative to the XAS $L_3$ peak energy(bottom x-axis).}
    \label{fig:img1}
\end{figure*}

For verifying the plasmon origin of the satellites, we measured another pair of core-levels, namely, the Te 3\textit{d} core levels, 
using HAXPES and SXPES as shown in Fig. 3(a) and 3(b). A least-squares fitting to the Te 3\textit{d}\textsubscript{5/2} and Te 3\textit{d}\textsubscript{3/2} main peaks and weak satellites is superimposed as full lines on the experimental spectra (empty circles). Here again, the main peaks were fitted with single asymmetric Voigt functions, and the satellites with symmetric Gaussian functions. The peak energy positions and peak FWHMs are listed in Table II.
The sharp intense lines at 572.53 eV and 582.94 eV in HAXPES and 572.50 eV and 582.89 eV in SXPES are the Te 3\textit{d}\textsubscript{5/2} and Te 3\textit{d}\textsubscript{3/2} main peaks, respectively,
Their observed BEs are very consistent with earlier reports of measured Te 3\textit{d} core levels of CoTe\textsubscript{2}\cite{muhler1998,hu2023}. The measurements reported by Hu et al.\cite{hu2023} was on samples from the same batch as present work, while the aim of that study was to check the feasibility of making terahertz nonlinear Hall rectifiers using mechanically exfoliated CoTe$_2$.
The narrow single main peaks of 3\textit{d}\textsubscript{5/2} and Te 3\textit{d}\textsubscript{3/2} and the absence of any feature $\approx$3.0 eV above the main peaks indicate absence of oxidation in the HAXPES and SXPES spectra\cite{hu2023}. Most importantly, the broader low intensity features are positioned at 19.65$\pm$0.1 eV higher BEs and confirm the plasmon origin, consistent with Te 3\textit{p} HAXPES and SXPES core levels. 
This hints at the possibility of a plasmon feature of the Co $2p_{3/2}$ main peak overlapping the satellite feature of the Co $2p_{1/2}$ main peak.

In order to check the above possibility, we then carried out a least square fit of the Co 2\textit{p} core levels of HAXPES and SXPES shown in Fig. 4(a) and 4(b), respectively. The fitting results showing the peak energy positions and peak FWHMs are listed in Table III. 
The Co 2\textit{p}\textsubscript{3/2} and Co 2\textit{p}\textsubscript{1/2} main peaks could be fitted with asymmetric Voigt line shapes, consistent with the metallic nature of CoTe\textsubscript{2}. Their observed BEs are 778.06 eV and 792.86 eV, respectively for HAXPES, and 777.90 eV and 792.70 eV, respectively for SXPES. These values are also consistent with previously measured Co 2\textit{p} core levels of CoTe\textsubscript{2}\cite{muhler1998, hu2023}. 
While the Co 2\textit{p}\textsubscript{3/2} range could be fitted by a single main peak and a single weak satellite feature, the Co 2\textit{p}\textsubscript{1/2} spectral range required a single main peak and two weak features.
From the values of the peak energy positions (Table III), it is clear that the second weak feature lies at a BE of 19.8$\pm$0.1 eV above the Co $2p_{3/2}$ main peak, consistent with the plasmons lying at a BE of 19.8$\pm$0.2 eV/19.65$\pm$0.1 eV above the Te $3p$ and $3d$ spectra, respectively. This confirms that the second weak feature corresponds to a plasmon feature of the Co $2p_{3/2}$ main peak, lying very close to the Co $2p_{1/2}$ satellite.
It was important to identify and separate out the plasmon feature from the Co $2p_{1/2}$ satellite, as it then allowed us
to carry out a cluster model calculation of the intrinsic main peak and satellites of the Co 2$p$ spectrum shown in Fig. 8(a) later.

\subsection{Co $2p$-$3d$ Resonant -PES of CoTe$_2$}

\par Figure 5(a) shows the Co 2\textit{p}-3\textit{d} resonant-PES valence band intensity map plotted as a function of incident photon energies ($h\nu$ = 770-803 eV) versus binding energy (BE = -1.2 to 45.8 eV). In order to obtain the resonant-PES map, we first measure the Co $L_3$- and $L_2$-edge XAS shown in Fig. 5(b) as a function of $h\nu$ (top x-axis). This provides us the $h\nu$ range to obtain resonant-PES valence band spectra which constitute the map. In order to clarify the map features, Fig. 5(c) shows valence band spectra (BE = -1.2 to 35.0 eV) measured at select $h\nu$ values (labelled $a$-$z$) across the $L_3$- and $L_2$-edges of Fig. 5(b). The BEs were calibrated with respect to $E_F$ of metallic CoTe\textsubscript{2} and the spectra are normalized to the shallow Te 4$d_{5/2, 3/2}$ core-level peaks (bright vertical lines at $\sim$40 and 42 eV BEs) in the Fig. 5(a) map and also for the spectra shown in Fig. 5c.
This was done to describe the resonance effects in the main Co 3d states close to h$\nu$ = $g$, and the evolution of the two-hole correlation satellite peak for $h\nu$ $>$ $g$. The map and Fig. 5(c) spectra show a small peak at 0.8 eV BE which gets enhanced on increasing $h\nu$ from $a$-$g$, and then gradually decreases for  $h\nu$ = $h$-$o$ across the $L_3$-edge. On further increasing $h\nu$ from $p$-$z$ across the $L_2$-edge, the 0.8 eV feature undergoes another maxima at $h\nu$ = $t$. This indicates that the 0.8 eV BE feature shows a Co 2\textit{p}-3\textit{d} resonance, thus identifying the single-particle Co $3d$ PDOS.  The 0.8 eV BE Co $3d$ PDOS feature  (blue vertical line in Fig 5(c)) is also seen as a narrow vertical bright line in the map. 
The main feature in the map is a high intensity diagonal (red full line) from the lower right to upper left corner. This high intensity diagonal originates from a peak feature at $\sim$4.5 eV BE which shows an intensity increase for $h\nu$ = $a$-$e$, marked as an orange dashed line in the map and Fig. 5(c). As can be clearly seen in Fig. 5(c), the  4.5 eV BE peak systematically moves to higher BEs with a shift equal to the increase in $h\nu$ and corresponds to the Co $L_{3}VV$ Auger feature. This is confirmed by plotting in Fig. 5(d) the kinetic energy of this peak as a function of $h\nu$ relative to the XAS $L_3$ peak energy (bottom x-axis). The actual incident $h\nu$ values are the same as top x-axis of Fig. 5(b). Thus, the peak at $\sim$4.5 eV BE is a resonant Raman feature and becomes the correlation satellite with two holes in the final state, as is known for elemental metals.\cite{Guillot,weinelt1997,hufner2000}. The map and Fig. 5(c) also show a weak broad feature at $\sim$19.5 eV BE above the correlation satellite, which is due to plasmon excitations, as confirmed by core-level spectra shown above. Further, for higher $h\nu$ = $p$-$z$, the map and Fig. 5(c) again show a weak resonance behavior for the Co $L_{2}VV$ resonant Raman (orange dashed line) and the two hole correlation satellite Auger peak (red full line) of the 4.5 eV BE feature. The corresponding kinetic energy of this peak as a function of $h\nu$ is also plotted in Fig. 5(d) and confirms its two-hole final state Auger character.

\begin{figure}
\includegraphics[scale=0.08]{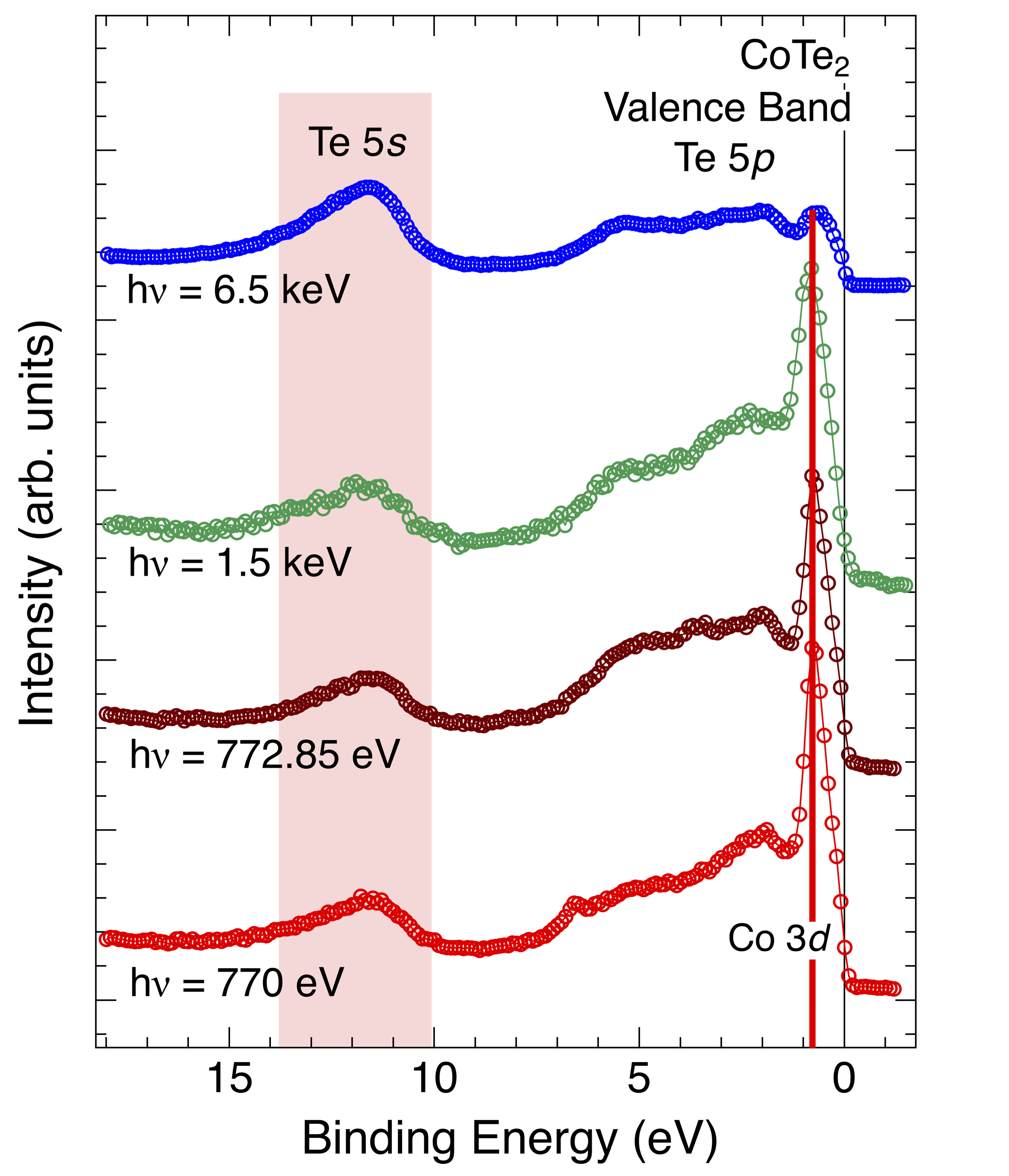}
\caption{CoTe\textsubscript{2} valence bands measured at different photon energies, namely h$\nu$ = 770 eV, h$\nu$ = 772.85 eV, h$\nu$ = 1.5 keV and h$\nu$ = 6.5 keV}
\end{figure}

\subsection{Off-resonant valence band spectra of CoTe$_2$}

With the aim of separating out the Te $5p$ partial density of states (PDOS) from the Co $3d$ PDOS based on their photoionization cross-sections (PICS)\cite{trzhaskovskaya2018}, we plot in Fig. 6 the valence band spectra over a large BE energy range using off-resonant photon energies $h\nu$ = 770 eV, 772.85 eV, 1.5 keV and 6.5 keV. The Fig. 6 spectra are normalized at the Te $5s$ shallow core level at $\approx$12 eV BE so as to emphasize the relative spectral  weights of Te $5p$ states and Co $3d$ states. Fig. 6 shows that the weight of Co $3d$ PDOS peak at 0.8 eV BE does not change much for $h\nu$ = 770 eV, 772.85 eV, and 1.5 keV  but gets strongly suppressed for $h\nu$ = 6.5 keV. This is due to the strongly reduced Co 3\textit{d} PICS compared to Te $5p$ PICS at $h\nu$ = 6.5 keV. The Te $5p$ states are spread from $E_F$ (as will be clarified below) to about 7 eV BE and show small variation in intensity due the change in PICS with $h\nu$ = 770 eV, 772.85 eV, 1.5 keV. But at $h\nu$ = 6.5 keV, the Te $5p$ states dominate the spectrum as the PICS\cite{trzhaskovskaya2018} ratio of Te $5p$:Co $3d$ is 9.47. Most importantly, the small peak near $E_F$ measured with $h\nu$ = 6.5 keV shows a different shape and width compared to the Co $3d$ peak with lower $h\nu$ values and this will discussed in more detail in Fig. 7(a) below, together with off- and on-resonant Co $2p-3d$ resonant-PES spectra. 

\begin{figure}
\includegraphics[scale=0.25]{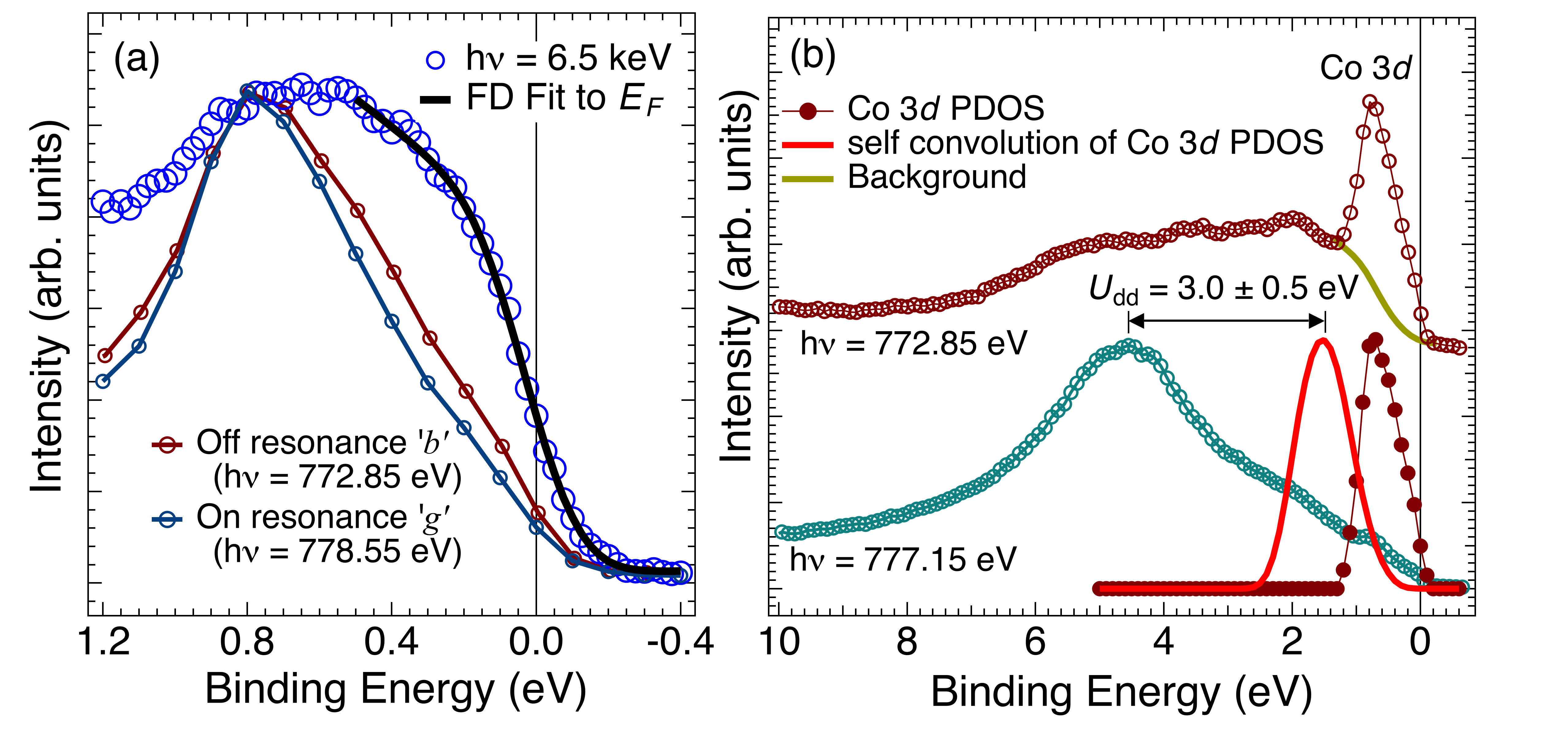}
\caption{(a) The CoTe$_2$ off-resonant ($h\nu$ = 772.85 eV, maroon, $\medcirc$; $h\nu$ = 6.5 keV, blue, $\medcirc$) and on-resonant ($h\nu$ = 778.55 eV, dark blue, $\medcirc$) near $E_F$ spectra (normalized at 0.8 eV BE). (b) Co 3\textit{d} PDOS (maroon, $\medbullet$) obtained by subtracting an integral background (gray line) from off-resonant spectrum($h\nu$ = 772.85 eV (maroon, $\medcirc$). The average $U_{dd}$ is the energy between self-convoluted Co $3d$ PDOS peak(red line) and the Resonant Raman peak ($h\nu$ = 777.15 eV; green, $\medcirc$), which becomes the $L_3VV$ Auger peak.}
\end{figure}

In order to identify the Te $5p$ PDOS at and near $E_F$, in Fig. 7(a) we plot  the off-resonant ($h\nu$ = 772.85 eV, maroon, $\medcirc$) and on-resonant ($h\nu$ = 778.55 eV, dark blue, $\medcirc$) near $E_F$ CoTe$_2$ spectra after normalizing their peak intensities at 0.8 eV BE, and compare them with the CoTe$_2$ spectrum measured with $h\nu$ = 6.5 keV (blue, $\medcirc$). At $h\nu$ = 6.5 keV, the Te $5p$ states dominate the spectrum as the photo-ionization cross-section\cite{trzhaskovskaya2018} ratio of Te $5p$:Co $3d$ is 9.47. The off- and on-resonant spectra show similar spectral shapes but the on-resonant spectrum gets narrowed, with suppressed relative spectral weight near $E_F$ without a clear Fermi edge. This indicates that the Co $3d$ PDOS peak  at 0.8 eV BE contributes weak spectral weight at $E_F$. In contrast, the CoTe$_2$ spectrum with $h\nu$ = 6.5 keV shows a broader peak compared to the Co $3d$ PDOS peak. It extends all the way upto $E_F$ and the leading edge matches the Fermi-Dirac (FD) fit, indicating that this feature is mainly derived from Te $5p$ states. This is consistent with ARPES experiments and comparison with DFT calculations which showed dominantly Te $5p$ states at and within 0.5 eV of $E_F$\cite{Chakraborty2023}. As we will show below (Fig. 8), this observation is also consistent with a negative-$\Delta$ in CoTe$_2$, but prior to that, we quantify $U_{dd}$ by applying the Cini-Sawatzky method\cite{cini1976, cini1977, sawatzky1977, cini1978} to the Co $3d$ PDOS and the two-hole correlation satellite data.

Figure 7(b) shows the off-resonant spectrum obtained with $h\nu$ = 772.85 eV before/after (maroon empty/full circles) subtracting an integral background (gray line) in order to separate out the single-particle Co $3d$ PDOS from the Te $5p$ states at higher BEs (see Fig. 6 and related discussion). The single-particle PDOS was then numerically self-convoluted to obtain the two hole spectrum (red line), and its peak energy represents the average two-hole energy without correlations. The two hole spectrum without correlations (red line) was then compared with the spectrum exhibiting the experimental two-hole correlation satellite spectrum in the resonant Raman region ($h\nu$ = 777.15 eV; green empty circles). The energy separation between the peak in the two-hole spectrum without correlations and the peak of the experimental two-hole correlation satellite gives $U_{dd}$ in the Cini-Sawatzky method. We obtain a value of $U_{dd}$ = 3.0 eV, indicating that CoTe$_2$ is a moderately correlated material. We then used the obtained $U_{dd}$ in CT cluster model calculations to simulate the Co $2p$ core-level PES and $L$-edge XAS in order to independently validate the $U_{dd}$ value.

\begin{figure}
\includegraphics[scale=0.10]{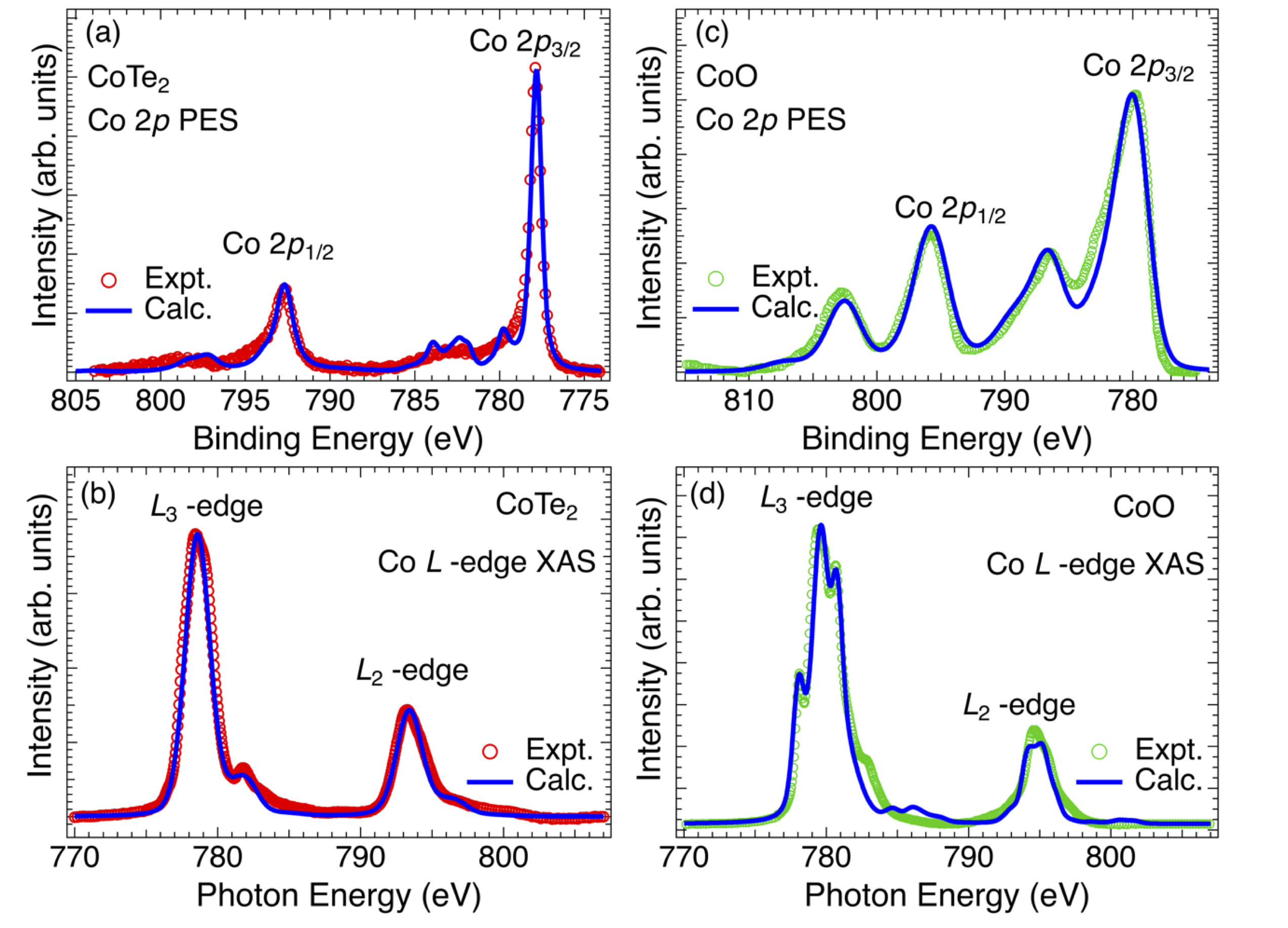}
\caption{(a) Co 2\textit{p} PES core levels and (b) Co \textit{L}\textsubscript{3,2}-edge XAS of CoTe\textsubscript{2}  compared with charge transfer cluster model calculations. (c) Co 2\textit{p} PES core levels and (d) Co \textit{L}\textsubscript{3,2}-edge of CoO  compared with charge transfer cluster model calculations. The Co $2p$ PES spectrum of CoO shown in panel (c) was taken from ref.\cite{AC} \copyright American Physical Society.}
\end{figure}

\subsection{Comparison of experimental and calculated Co 2p PES and Co L-edge XAS spectra of CoTe$_2$ and CoO}

Figure 8(a,b) and (c,d) show experimentally measured Co $2p$ core-level PES and $L$-edge XAS spectra of CoTe$_2$ and CoO, respectively. A weak plasmon feature partly overlapping the Co $2p_{1/2}$ satellite of the CoTe$_2$ spectrum was removed after a fitting procedure(see Fig. 4). The Co $2p$ PES spectrum of CoO was taken from our earlier work\cite{AC}. The corresponding calculated spectra (full lines) are also overlaid on the experimental spectra (symbols).  The calculated spectra were obtained from CT atomic multiplet cluster model calculations\cite{deGroot,GrootKotani}. We used the QUANTY code\cite{haverkort2012, lu2014, haverkort2014} to calculate the spectra as described in detail in the Methods section.  We used a Co$L_6$ cluster (where, $L$ is the ligand tellurium/oxygen atom) in an octahedral (O$_h$) local symmetry, as is known from the crystal structures of CoTe$_2$\cite{Chakraborty2023}  and CoO\cite{Johnston}. The calculations require the metal ion valency as an input. It is known\cite{Jobic} that CoTe$_2$ exhibits a stable divalent state of Co$^{2+}$ with Te atoms in a dimerized (Te$_2^{2-}$) configuration with a reduced Te-Te distance along the c-axis. This is a generic feature of several late $3d$, $4d$ and $5d$ TM tellurides crystallizing in the so called polymeric 1$T$-CdI$_2$ structure\cite{Jobic, bensch1996}. In a recent study of nanomaterial CoTe$_2$/Carbon nanotube hybrid material, as well as pure CoTe$_2$ nanowires(70 nm width), it was shown that the Co $L_3$-edge exhibits a Co$^{2+}$ narrow main peak at $h\nu$= 778.8 eV\cite{lu2016}.
Our data of Fig. 8(b) also shows a very similar spectrum with the Co $L_3$ main peak at $h\nu$ = 778.8 eV and confirms the divalent Co$^{2+}$($d^7$) configuration in CoTe$_2$. 
Since XAS is a local probe which measures site- and orbital-projected states (due to dipole-transition selection rules)\cite{deGroot,GrootKotani}, the XAS spectra are very similar for bulk CoTe$_2$ and nanomaterial CoTe$_2$. This is further confirmed by the fact that we can consistently calculate the Co $L$-edge XAS spectra using an octahedral cluster model, which is a well-established method to analyze XAS spectra\cite{deGroot,GrootKotani}.
On the other hand, the Co $L_3$-edge XAS of Co$^{2+}$ in CoO is known to exhibit a broad structured peak with the lowest multiplet prepeak at $h\nu$ = 778.1 eV, a main peak multiplet at $h\nu$ = 780.0 eV with a shoulder at $h\nu$ =781 eV \cite{Elp,Okamoto}, and our data is very consistent with these studies. For the Co $2p$ PES, the $2p_{3/2}$ main peak is at 777.9 eV for CoTe$_2$  and at 780.0 eV for CoO, consistent with reported values\cite{hu2023,Elp,Ghiasi,XPShandbook}.

Based on the above, the Co$^{2+}$ initial state for CoTe$_2$ and CoO is considered to be a linear combination of three basis states: $3d^7$, $3d^{8}\underline{L}^1$ and $3d^{9}\underline{L}^2$. The calculation was carried out for an octahedral ML$_6$ cluster as explained in Methods section. 
While CoTe$_2$ forms in the 1T-CdI$_2$ structure, and CoO forms in the rock salt structure, they both share a similar local octahedral symmetry. It is noted that CoTe$_2$ actually exhibits a distortion from octahedral symmetry, but since the distortion is very small, CoTe$_2$ results can be compared with CoO results due to the local octahedral symmetry of CoO.
Using the experimentally obtained $U_{dd}$ = 3.0 eV for CoTe$_2$ and a value of $U_{dd}$ = 5.0 eV for CoO (from ref. \cite{Elp}), we carried out an extensive set of calculations varying the values of 
$\Delta$, $T_{eg}$, $T_{t2g}$ (=$T_{eg}/2$) and $10Dq$ to obtain calculated spectra very similar to the experimental spectra, as shown in Figs. 8(a-d).
The same parameter set was used for calculating Co $2p$ core-level PES and $L$-edge XAS spectra of each material, and the main parameters are listed in Table IV.   The obtained parameters show that CoTe\textsubscript{2} is a negative-$\Delta$ system, while CoO is confirmed to be a positive-$\Delta$ system\cite{Ghiasi}. While this is the first analysis of the CoTe$_2$ spectrum, the parameters for CoO are quite close to earlier analyses using a cluster model\cite{Elp}, as well as a CT multiplet calculation combined with DMFT method\cite{Ghiasi}(see SM Table I for comparison of CoO parameters). The values of negative-$\Delta$, $U_{dd}$, as well as the Slater parameters F$_k$ and G$_k$ for CoTe$_2$ and CoO were checked by systematic calculations to determine optimal parameters as detailed in SM;SN1, Figs. S1-S2. A negative-$\Delta$ is expected\cite{Pavarini} by the trend in the reduction of electronegativity from 
 O$\rightarrow$S$\rightarrow$Se$\rightarrow$Te in column 6A of the periodic table.

\begin{table}[t!]
	
	 \caption{Electronic parameters and $d^n$ for materials, obtained using  3-basis state cluster model calculations.$^\dagger$from ref.\cite{shelke2025}; *from ref.\cite{Mizokawa}}
\begin{tabular}{ccccccc}

\hline
Parameter& ~CoTe$_2$~&~CoO~&~CrTe$^\dagger$~&~NaCuO$_2^*$~\\
\hline
$U_{dd}$(eV)		&~3.0~&~5.0~&~3.5~&~7.0~ \\
$\Delta$(eV)		&~-2.0~&~4.0~&~-1.0~&~-2.0~ \\

$T_{eg}$(eV)		&~1.2~&~2.5~&~1.3~&~2.7~ \\

$d^n$ count		&~8.14~&~7.21~&~4.98~&~8.81~ \\

$U_{dd}/T_{eg}$	&~2.5~&~2.0~&~2.7~&~2.6~ \\

$\Delta/T_{eg}$&~-1.7~&~1.6~&~-0.8~&~-0.7~ \\
\hline
\end{tabular}

\end{table}

We have calculated the  $d$-electron count in the ground state and the obtained values are 8.14 electrons for 
CoTe$_2$ (Table V). This indicates a dominant $3d^{n+1}\underline{L}^1$ contribution in the ground state (see Table VI) and is a signature of negative-$\Delta$ materials, as reported for insulating NaCuO$_2$\cite{Mizokawa}, metallic phase of the charge-density wave(CDW) rare-earth nickelates RNiO$_3$ which show metal-insulator transitions \cite{Bisogni,MizokawaPNO,Piamontze,Meyers,Green}, ferromagnetic metal CrTe\cite{shelke2025}, etc. It is noted that early cluster model calculations\cite{MizokawaPNO,Piamontze,Meyers} for the XAS of metallic RNiO$_3$ used a small positive-$\Delta$ while the most recent study\cite{Green} used a negative-$\Delta$,  but all of them concluded a dominantly charge-transferred ground state.

\subsection{Comparison of CoTe$_2$ and RNiO$_3$ electronic structure}

It is important to discuss and compare the electronic structure of RNiO$_3$ materials in the high-$T$ metallic phase with CoTe$_2$, as the formally Ni$^{3+}$ ions in RNiO$_3$ have the same $3d^7$ electron configuration like Co$^{2+}$ in CoTe$_2$. 
Early single metal-site cluster model calculations\cite{MizokawaPNO,Piamontze,Meyers} for the XAS of the metallic RNiO$_3$ phase used a small positive-$\Delta$ while the most recent study\cite{Green} used a negative-$\Delta$,  but all of them concluded a dominantly charge transferred ground state (with weights of $d^{n+1}\underline{L}^1$+$d^{n+2}\underline{L}^2$ $>$ $d^n$). In Table V, we compare the electronic parameters, the spin magnetic moment $m_S$, ground state weights and total $d^n$ counts for CoTe$_2$ with two cases of RNiO$_3$ reported in literature\cite{MizokawaPNO,Green}. It shows a dominantly $d^{n+1}\underline{L}^1$ ground state character using single metal-site cluster model calculations for RNiO$_3$, as obtained for CoTe$_2$ in present study.

\begin{table}[t!]
	
	 \caption{Electronic parameters, spin magnetic moments $m_S$, weights in the ground state and total $d^n$-counts for CoTe$_2$ and RNiO$_3$ (metal phase) using  single metal-site cluster model calculations.}
\begin{tabular}{cccc}

\hline
~~~~& ~~~CoTe$_2$~~~&~~~RNiO$_3$~~~&~~~RNiO$_3$~~~\\
~~~~&~~~~&~ref.\cite{MizokawaPNO}~&~ref.\cite{Green}~\\
Parameter        &~~~~&~~~~\\
\hline
$U_{dd}$(eV)		&~3.0~&~7.0~&~6.0~ \\
$\Delta$(eV)		&~-2.0~&~1.0~&~-0.5~\\

$T_{eg}$(eV)		&~1.2~&~2.6~&~3.0~\\
$T_{t2g}$(eV)		&~0.6~&~1.2~&~1.74~\\

$10Dq$(eV)		&~1.0~&~0.6~&~0.95~\\

$m_S$($\mu_B$)	&~0.9~&~0.9~&~1.1~\\

$d^7$ weight		&~11.0\%~&~34\%~&~24\%~\\

$d^{8}\underline{L}^1$ weight		&~64\%~&~56\%~&~61\%~\\

$d^{9}\underline{L}^2$ weight		&~25\%~&~10\%~&~15\%~\\

$d^n$ count	&~8.14~&~7.76~&~7.83~\\

\hline
\end{tabular}
	
\end{table}

Further, several theoretical calculations\cite{Mazin,Park,SJohnston,Subedi,Seth} have been carried out to address the CDW transition mechanism and role of Coulomb correlations. In spite of different theoretical methods, all of them agree that the Ni$^{3+}$ $3d^7$ state gets stabilized to a dominantly $3d^{8}\underline{L}^1$ ground state in the high-$T$ metallic phase. This is consistent with a negative-$\Delta$. Based on a low-energy description involving $e_g$-orbitals and LDA+$U$ calculations ($U$ = 5 eV; Hund's coupling $J_H$ = 1 eV), Mazin et al.\cite{Mazin} clarified the type of CDW. They showed a CDW of the type 2$e_g^1\rightarrow e_g^0+e_g^2$ is favored if $e_g$ bandwidth becomes larger than the Jahn-Teller splitting and $J_H$ reduces $U$ to $U_{eff}$. Park et al.\cite{Park} carried out DFT+DMFT calculations (also with $U$ = 5 eV; $J_H$ = 1 eV) and showed that neighboring Ni sites show long Ni$_1$-O bonds ($3d^{8}$ with S=1; paramagnetic Curie-type local susceptibility, $\chi$($T$)$\sim$1/$T$) and short Ni$_2$-O bonds($3d^{8}\underline{L}^2$ with S= 0;  paramagnetic metal type $T$-independent $\chi$). They called it a ``site-selective Mott transition". Johnston et al.\cite{SJohnston} reported exact diagonalizaton calculations of Ni$_2$O$_{10}$ clusters as well as Hartree-Fock calculations. They obtained a metal to a CDW gapped state with two types of Ni ions, upon increasing distortion. Subedi et al.\cite{Subedi} calculated a phase diagram using DFT+DMFT calculations. They showed that if  $U_{eff}$=$U-3J_H$ $\lesssim$ 0, where $U$ and $J_H$ are not the atomic values but the renormalized values for hybridized $e_g$ states, it causes a spontaneous bond disproportionation for large enough $J_H$. Seth et al.\cite{Seth} used a GW+DMFT scheme\cite{Biermann} to show that $U$ and $J_H$ indeed get reduced in RNiO$_3$. Further, a CDW phase is stabilized on including an intersite Coulomb interaction. These studies\cite{Mazin,Park,SJohnston,Subedi,Seth} concluded that a small or negative $U_{eff}$ is qualitatively consistent with a negative-$\Delta$. 
However, our results for determining $U_{dd}$ using the  Cini-Sawatzky analyses indicate moderate values of $U_{dd}$ = 3.0 eV for CoTe$_2$. Nonetheless, using these values of  $U_{dd}$, our calculations 
indicate CoTe$_2$ is a negative-$\Delta$ material with a dominant $3d^{n+1}\underline{L}^1$ contribution in the ground state. This indicates a very similar electronic structure like the high temperature metallic phase of RNiO$_3$.

While CoTe$_2$ is not a CDW system, the present Cini-Sawatzky analysis results indicate a moderate value of $U_{dd}$ = 3.0 eV for CoTe$_2$ although it can actually provide even negative-$U_{dd}$ values, as was shown for TiSe$_2$ and CrSe$_2$\cite{deBoer,Chuang}. 
We have also checked that the occupied $3d$-bands\cite{Chakraborty2023} have a width $W_d$$<$$U_{dd}$  and hence, the cluster model is  applicable to analyze CoTe$_2$.
Table IV also lists known parameters from two other negative-$\Delta$ materials, CrTe\cite{shelke2025} and NaCuO$_2$\cite{Mizokawa}, and all show a dominantly $3d^{n+1}\underline{L}^1$ ground state. 
 Interestingly, Table IV shows that for negative-$\Delta$ cases, while $U_{dd}$ varies between 3.0 eV to 7.0 eV, the scaled Coulomb energy $U_{dd}/T_{eg}$ and $\Delta/T_{eg}$ show small variation. It indicates that all the negative-$\Delta$ materials in Table IV lie in a small region in the ZSA phase diagram.
From Table IV, it is clear that the $d$-$p$ hybridization strength $T_{eg}$ for CoTe$_2$$<$CoO, related to the fact that the average Co-Te distance (=2.565$\AA$)\cite{Chakraborty2023}  is much larger than the average Co-O distance (=2.13$\AA$)\cite{Johnston}. This suggests that the reduction of $U_{dd}$ in CoTe$_2$ compared to CoO is not caused by $T_{eg}$.

\begin{figure}
\includegraphics[scale=0.265]{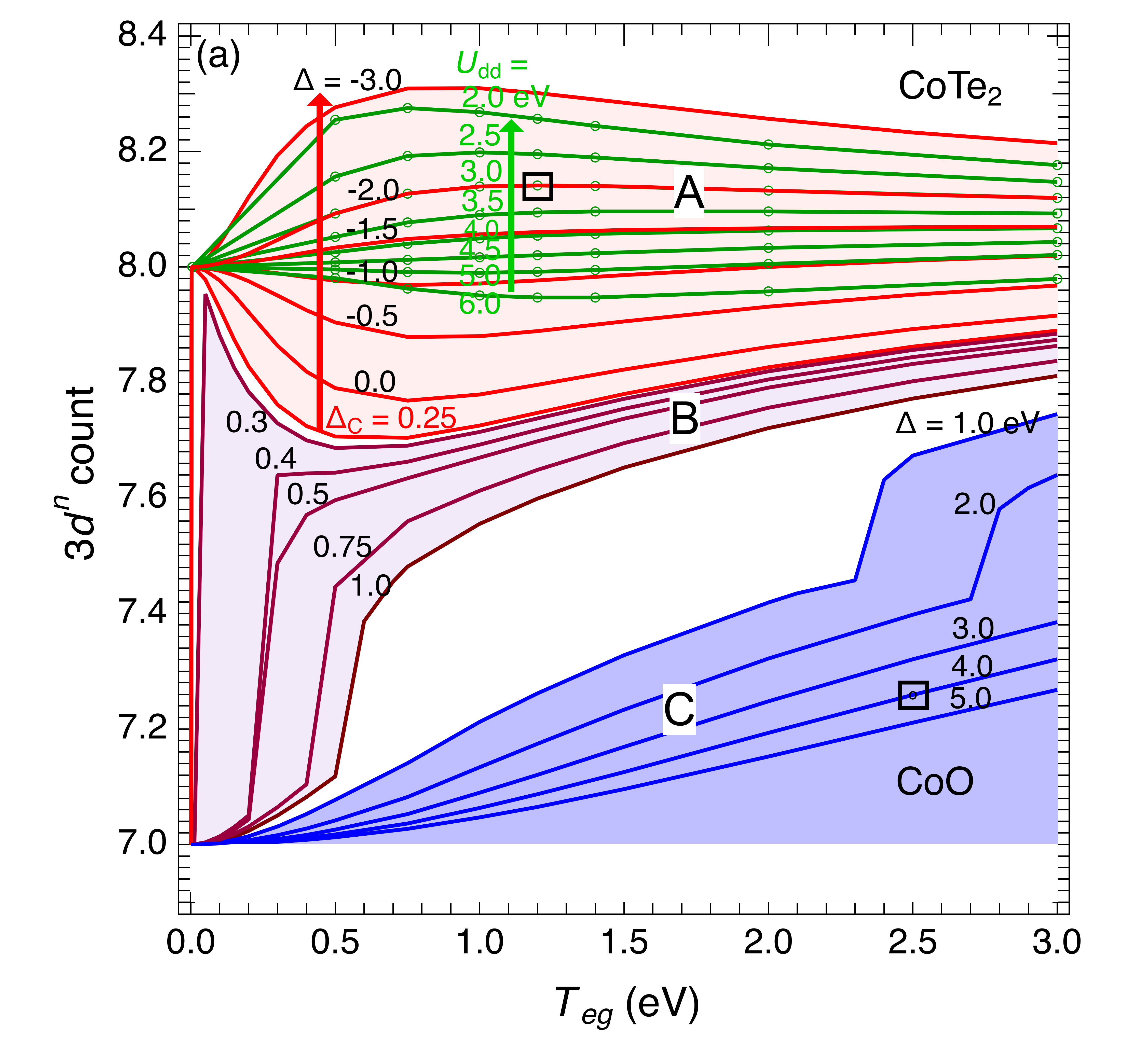}
\caption{Plots of $d^n$-count vs. $T_{eg}$ for selected values of $\Delta$ and $U_{dd}$ of CoTe$_2$ and CoO, identify regions of effective negative-$\Delta$ (A) and effective positive-$\Delta$ (B, C). Squares($\square$) indicate optimal values which reproduce experimental spectra (Fig. 8).}
\end{figure}

\subsection{Characterizing the effective negative-$\Delta$($\leq$$\Delta_C$) and effective positive-$\Delta$($>$$\Delta_C$) regions}

Next, we checked the relation of $\Delta$ and $U_{dd}$ with $d^n$ using CT cluster model calculations. Fig. 9 (red curves) shows the variation of $d^n$ count vs. $T_{eg}$ for different $\Delta$ values (keeping all other parameters fixed to optimal values for CoTe$_2$). The plots show that for Co$^{2+}$, starting with $d^n$ = 7 for $T_{eg}$ = 0, we obtain a sharp increase of $\sim$1 electron to $d^n$$\sim$8 for the smallest considered finite $T_{eg}$ = 5 meV, for all negative $\Delta$, and also upto a small positive $\Delta$$\leq$$\Delta_C$ = 0.25 eV. This indicates a spontaneous charge transfer takes place due to an effective negative-$\Delta$ to form the dominantly $3d^{n+1}\underline{L}^1$ ground state for small $T_{eg}$ = 5 meV. For optimal parameters ($\Delta$= -2.0 eV and $T_{eg}$ = 1.2 eV) corresponding to spectra shown in Figs. 8(a,b), we obtain  $d^n$ = 8.14 (black square in red curve for $\Delta$= -2.0 eV in region A of Fig. 9). For $T_{eg}$ = 5 meV(see following discussion, Fig. 10(a)), the spin magnetic moment $m_S$ vs. $\Delta$ also exhibits a jump at $\Delta_C$ , while for optimal $T_{eg}$ = 1.2 eV,  $m_S$ vs. $\Delta$ does not show a jump but a smooth variation across $\Delta_C$.
We then checked the variation of $d^n$ count vs. $T_{eg}$ for $U_{dd}$ = 2.0 to 6.0 eV (green curves; all other parameters fixed to optimal values for CoTe$_2$). The results in Fig. 9 show that in region A (pink shade), an increase in $d^n$ is obtained on reducing $\Delta$ (making it more negative;red arrow) or reducing $U_{dd}$(green arrow). This indicates that the relatively large CT from ligand to Co site in CoTe$_2$ originates from a combination of effective negative-$\Delta$$\leq$$\Delta_C$ and reduced $U_{dd}$. The microscopic origin of the reduction of $U_{dd}$ is the polarizability of the anions, which is larger 
for Te than oxygen anions since the polarizability is roughly proportional to the anion size\cite{ZS}. Further, for $\Delta$$>$$\Delta_C$ (region B; purple shade), the sharp jump in $d^n$ for $T_{eg}$ = 5 meV gets suppressed for higher $T_{eg}$ in the form of a reduced jump, followed by a gradual change at higher $T_{eg}$.

In the following, we clarify several aspects of the properties of the effective negative and positive charge-transfer energy $\Delta$ regions labelled A-C in Fig. 9 across the critical $\Delta_C$ in CoTe$_2$. Figs. 9 show the evolution of the total electron count $d^n$ as a function of $T_{eg}$ for various values of the $\Delta$ and $U_{dd}$,
and these energies are defined as multiplet averaged values\cite{Bocquet,Fujimori,Fujimori2}.
We then address the role of an effective negative $\Delta$ compared to the multiplet averaged $\Delta$ and why do we obtain $\Delta_C$ = 0.25 eV in Fig. 9 when we use other parameters fixed to CoTe$_2$ optimal parameters.

\begin{figure*}
\centering
\includegraphics[scale=0.25]{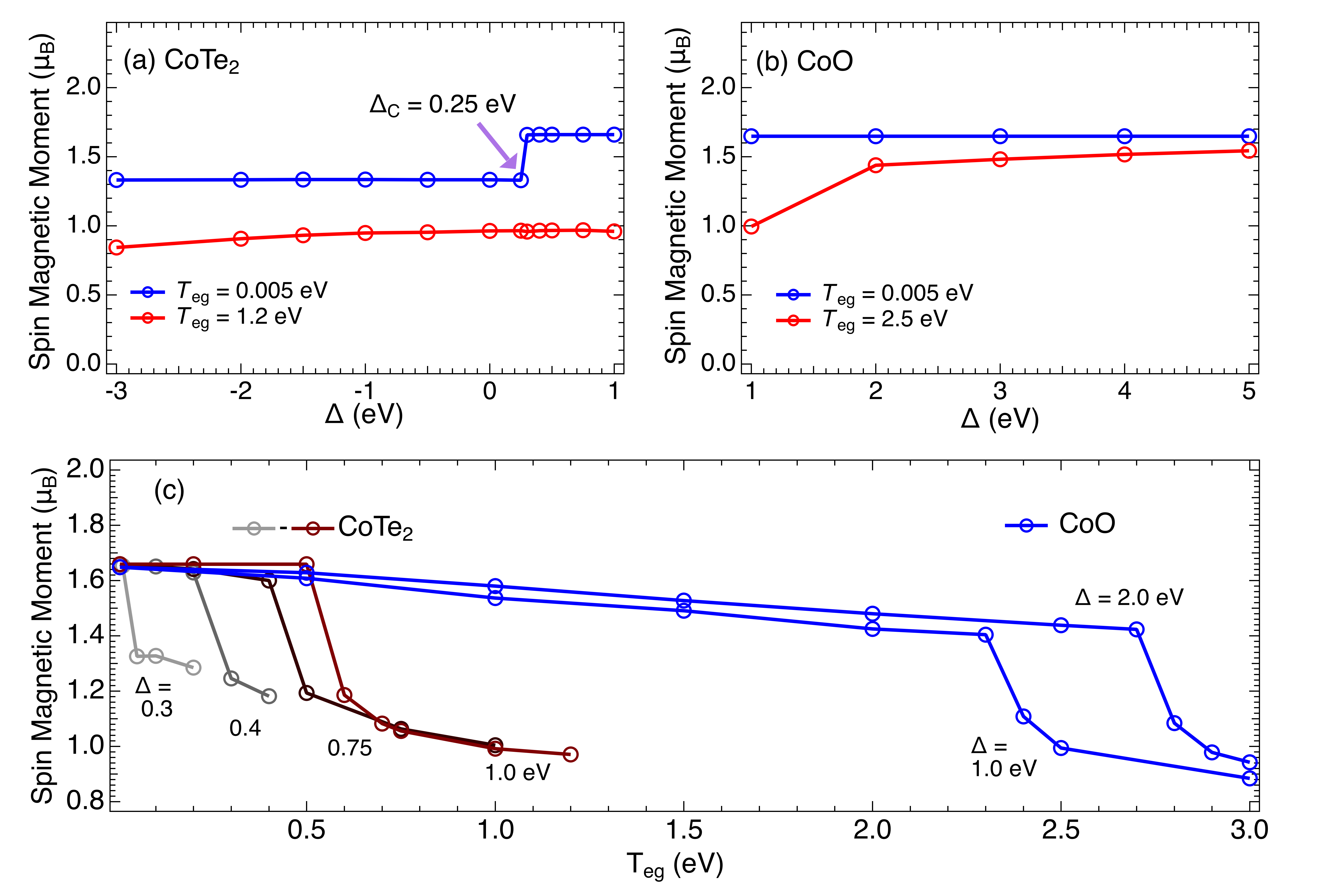}
\caption{(a) In CoTe$_2$, the spin magnetic moment $m_S$ vs. $\Delta$ for $T_{eg}$ = 5 meV, exhibits a jump at $\Delta_C$ = 0.25 eV, while for optimal $T_{eg}$ = 1.2 eV,  $m_S$ vs. $\Delta$ shows a gradual change across $\Delta_C$.
 (b) In CoO, the $m_s$ vs. $\Delta$ shows a nearly constant value for small $T_{eg}$ = 5 meV, while for $\Delta$$\leq$2.0 eV, a small jump is observed in $m_S$ vs. $T_{eg}$, for optimal $T_{eg}$ = 2.5 eV. (c) The jumps seen in regions B and C of Fig. 9 are also confirmed to be associated with spin-state transitions.}
\end{figure*}

\begin{table}[t!]
	
	 \caption{Ground state weights (\%) obtained from cluster model calculations with $T_{eg}$ = 5 meV, starting with the formal $d^7$ configuration for CoTe$_2$}
\begin{tabular}{cccc}

\hline

CoTe$_2$					&$d^7$&$d^{8}\underline{L}^1$&$d^{9}\underline{L}^2$\\
\hline 
$\Delta$ = $\Delta_C$ = 0.25 eV		&~0.1\%~&~99.9\%~&~0.0\%~	\\
(and all $\Delta$$<\Delta_C$) 				&		&		&			\\
\hline
$\Delta$ =	 0.3 eV			&~99.9\%~&~0.1\%~&~0.0\%~			\\
(and all $\Delta$$>\Delta_C$)				&		&		&			\\
\hline \\

\end{tabular}

\end{table}

In Fig. 10(a), we plot the spin magnetic moment $m_S$ vs. $\Delta$ for $T_{eg}$ = 5 meV obtained from the 
same cluster model calculation results shown in Fig. 8(a), using other parameters fixed to CoTe$_2$ optimal parameters. The $m_S$ values exhibit a jump at $\Delta$ = 0.25 eV, where even the total $d^n$ count shows a jump in Fig. 8(a), and we denote it as $\Delta_C$. The $m_S$ values show negligible change for $\Delta$$\leq$$\Delta_C$, and also for $\Delta$$>$$\Delta_C$. 
On the other hand, $m_S$ vs. $\Delta$ curve for the CoTe$_2$ optimal value of $T_{eg}$ = 1.2 eV, the jump in $m_S$ values gets suppressed and it shows a small gradual increase on increasing $\Delta$.
The corresponding individual ground state weights from the cluster model calculations for $T_{eg}$ = 5 meV are listed in Table VI.
It is clear from Fig. 10(a) and Table VI that for $\Delta$$\leq$$\Delta_C$, the results indicate a dominantly
 $d^{n+1}\underline{L}^1$ state corresponds to the effective negative-$\Delta$ region A of Fig. 9, and for 
$\Delta$$>$$\Delta_C$, it indicates a dominantly  $d^{n}$ state and corresponds to the effective positive-$\Delta$ region B of Fig. 9.

Fig. 10(b) shows the spin magnetic moment $m_S$ vs. $\Delta$ for $T_{eg}$ = 5 meV obtained from the cluster model calculation results of CoO shown in Fig. 8(c,d), using other parameters fixed to CoO optimal parameters. The $m_S$ values show a negligible change for all $\Delta$ values, while the $m_S$ vs. $\Delta$ curve for the optimal value of $T_{eg}$ = 2.5 eV shows a small gradual increase of $m_S$ on increasing $\Delta$. The results indicate that region C of Fig. 9 corresponds to the usual positive-$\Delta$ region. 
For optimal parameters ($\Delta$= 4.0 eV and $T_{eg}$ = 2.5 eV) corresponding to spectra shown in Figs. 8(c,d), we obtain  $d^n$ = 7.21
(black square in blue curve for $\Delta$= 4.0 eV in region C of Fig. 9)
It is noted that the $m_S$ values on increasing $\Delta$ in Fig. 10(a) connect to the $m_S$ values on increasing $\Delta$ in Fig. 10(b), confirming that region B is an effective positive-$\Delta$ region. As another check,
in Fig. 10(c), we plot the $m_S$ vs. $T_{eg}$ for selected $\Delta$ values from regions B and C of Fig. 9. The results show jumps corresponding to those seen in the total $d^n$ count in Fig. 9, and confirm they are associated with spin-state transitions. The corresponding individual ground state weights from the cluster model calculations with $T_{eg}$ = 5 meV are also listed in Table VI.

We then answer the question about why $\Delta_C$ = 0.25 eV in CoTe$_2$. The  $\Delta$ as used in the present cluster model calculations is defined as,

$\Delta$ = $E$($d^{n+1}\underline{L}^1$)- $E$($d^n$), 

where $E$($d^n$) is the center of gravity or average energy of the $d^n$ multiplets and $E$($d^{n+1}\underline{L}^1$) is the center of gravity or average energy of $d^{n+1}\underline{L}^1$ multiplets. 
The charge transfer energy can be also defined by using the energies of the lowest multiplet of $d^{n+1}\underline{L}^1$ and $d^n$  configurations\cite{Bocquet,Fujimori,Fujimori2}. This is relevant because the low energy properties of transition metal compounds are mainly determined by excitations associated with the lowest energy multiplets. If we denote the energy difference between 
the energy of the lowest multiplet and the average multiplet energy as $\Delta'$$E_n$, then the effective charge transfer energy
 $\Delta_{eff}$, is given by 

$\Delta_{eff}$ = $\Delta$ + $\Delta'$$E_{n+1}$ - $\Delta'$$E_n$,

Accordingly, one can check when does the ground state transform from a dominantly $d^{n}$ state to $d^{n+1}\underline{L}^1$ state i.e. negative $\Delta$ represents $\Delta$$<0$ or 
$\Delta_{eff}$$<0$ ?
Using the same cluster model calculations discussed above, we 
calculate $L$-edge XAS spectra with a small $T_{eg}$ = 5 meV for various $\Delta$ values at and across $\Delta_C$, with a very small Gaussian broadening  of 0.1 eV FWHM as shown in Fig. 11.
The $L$-edge XAS spectrum corresponds to transitions from the ground state to final states of the type
$2p^63d^n$$\rightarrow$$2p^53d^{n+1}$, $2p^63d^{n+1}\underline{L}^1$$\rightarrow$$2p^53d^{n+2}\underline{L}^1$, etc. From a careful check of the XAS spectrum, we try to decipher the role of the electronic parameters on the initial state multiplets of the $d^{n}$ and $d^{n+1}\underline{L}^1$ states.
In the following, we show that 
the actual transition to a negative charge-transfer character takes place when $\Delta_{eff}$$<0$, and 
$d^{n+1}\underline{L}^1$ contribution dominates the ground state.

\begin{figure}
\includegraphics[scale=0.8]{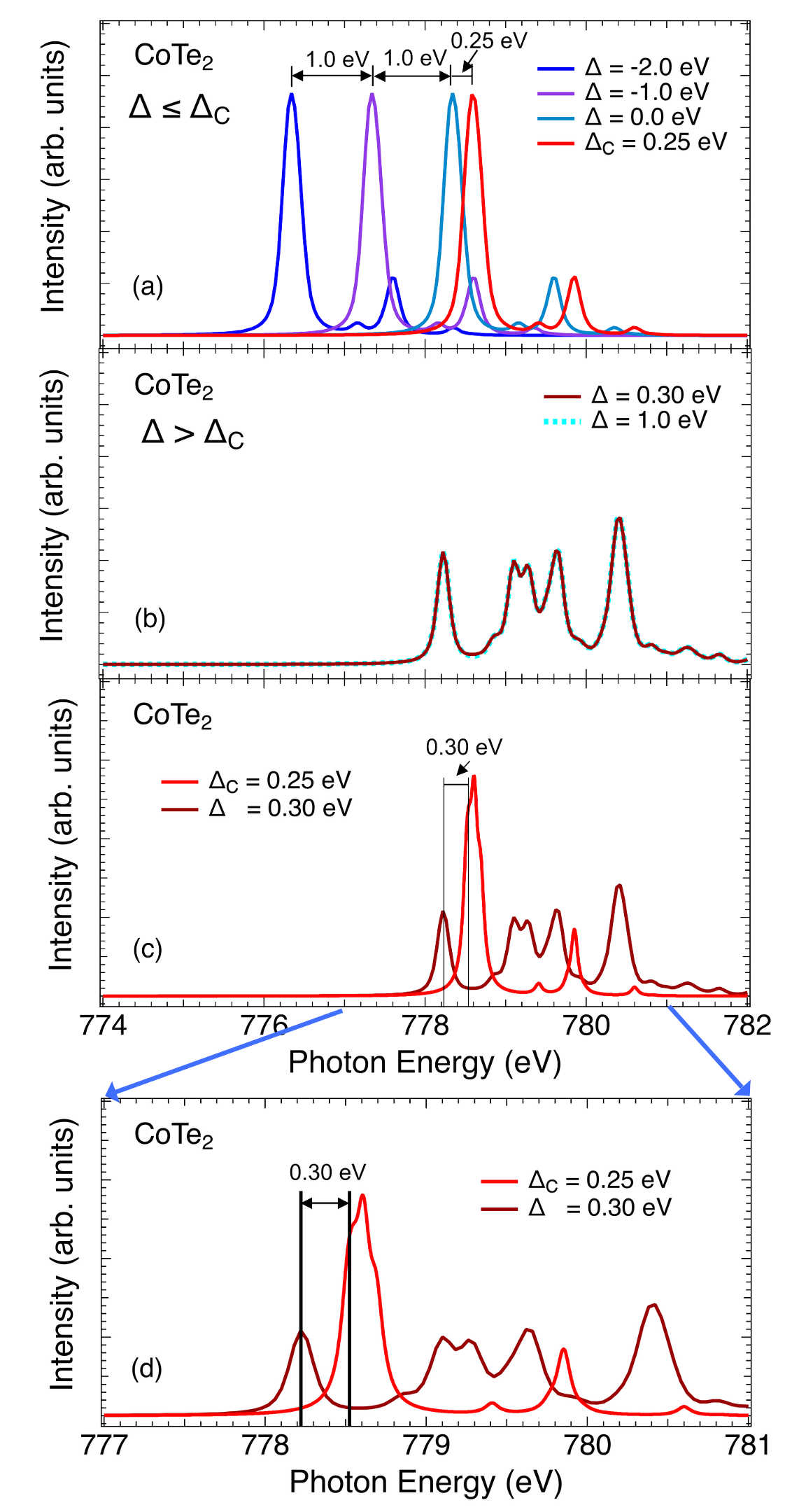}
\caption{ (a-d) The XAS spectral calculations for $T_{eg}$ = 5 meV and varying $\Delta$ with all other parameters fixed to optimal values of CoTe$_2$: (a) for $\Delta$$\leq$$\Delta_C$ = 0.25 eV, (b) for $\Delta$$>$$\Delta_C$ = 0.25 eV. (c) for $\Delta$ = $\Delta_C$ = 0.25 eV ($d^{n+1}\underline{L}^1$ state) and $\Delta$ = 0.30 eV ($d^{n}$ state) (d) same as (c), but plotted on an expanded x-scale.}
\end{figure}

Fig. 11(a-c) shows the $L$-edge XAS spectra with a small $T_{eg}$ = 5 meV and a small Gaussian broadening  of 0.1 eV FWHM for selected $\Delta$ values, with all other parameters fixed to optimal values of CoTe$_2$
The spectra in panel(a) for $\Delta$$\leq$$\Delta_C$ = 0.25 eV  show hardly any change in the shape of the multiplet features but do show a systematic shift equal to the change in $\Delta$. From Table VI, it is clear that all the spectra  originate in the dominantly $d^{n+1}\underline{L}^1$ initial state multiplets.
In panel (b), we plot two spectra for $\Delta$$>$$\Delta_C$ = 0.25 eV and the results show very similar spectra
but with significantly different multiplet features.
From Table VI, we know that for $\Delta$ = 0.30 eV, the initial state is dominated by the $d^{n}$ state. This indicates the spectra in panel (b) originate in the $d^{n}$ state and do not show a shift as is seen in Fig. 11(a) for $\Delta$$\leq$$\Delta_C$. 
In panel(c), we compare the spectra for $\Delta_C$ = 0.25 eV with $\Delta$ = 0.30 eV, and the spectrum $\Delta$ = 0.30 eV calculated with 0.01 eV Gaussian broadening to see fine features. Panel (d) shows the same spectra plotted on an expanded x-scale. The lowest energy multiplet of the $\Delta$ = 0.30 eV spectrum is shifted 0.3$\pm$0.1 eV below the lowest energy multiplet of the $\Delta_C$ spectrum. 
This difference of the spectral behavior for $\Delta$$\leq$$\Delta_C$ and $\Delta$$>$$\Delta_C$, together with the results of $d^n$ count of Fig. 9 indicate that region A with $\Delta$$\leq$$\Delta_C$ = 0.25 eV corresponds to the effective negative-$\Delta$ region, and region B with $\Delta$$>$$\Delta_C$ corresponds to an effective positive-$\Delta$ region.
In a recent study, we have investigated another Dirac semimetal NiTe$_2$, which forms in the same structure as CoTe$_2$. Using the same experimental techniques and cluster model calculations, it was found that NiTe$_2$ exhibits a $\Delta_C$ = -1.55 eV and shows a qualitatively similar phase evolution in a plot of $d^n$-count vs. $T_{eg}$ for a relevant set of values of $\Delta$ and $U_{dd}$ of CoTe$_2$\cite{ShelkeNiTe2}.

For CoO, a similar plot (Fig. 9; region C; blue shade) starts with  $d^n$ = 7 for $T_{eg}$ = 0, and shows a gradual increase with $T_{eg}$ for typical values of positive-$\Delta$. For small $T_{eg}$ = 5 meV, $m_s$ vs. $\Delta$ shows a nearly constant value (see Fig. 10(b)). However, for $\Delta$$\leq$2.0 eV, a small jump is observed in $d^n$ vs. $T_{eg}$, at high $T_{eg}$ values. We have checked that the reduced jumps in regions B and C are due to spin-state transitions, as expected from cluster model calculations(Fig. 10(c)). 

We have thus confirmed that $\Delta_C$ for CoTe$_2$ corresponds to attaining an effective negative $\Delta$, defined as the energy difference between lowest multiplet of $d^n$ and $d^{n+1}\underline{L}^1$  states\cite{Bocquet,Fujimori,Fujimori2}, i.e. a material attains a genuine negative-$\Delta$ state for $\Delta$$\leq$$\Delta_C$, when lowest multiplet of the $d^{n+1}\underline{L}^1$ state becomes more negative than lowest multiplet of the $d^{n}$ state. Fig. 9 highlights regions of effective negative-$\Delta$ (A) and effective positive-$\Delta$ (B, C). 

\subsection{Summary discussion}

\begin{figure}
\includegraphics[scale=0.12]{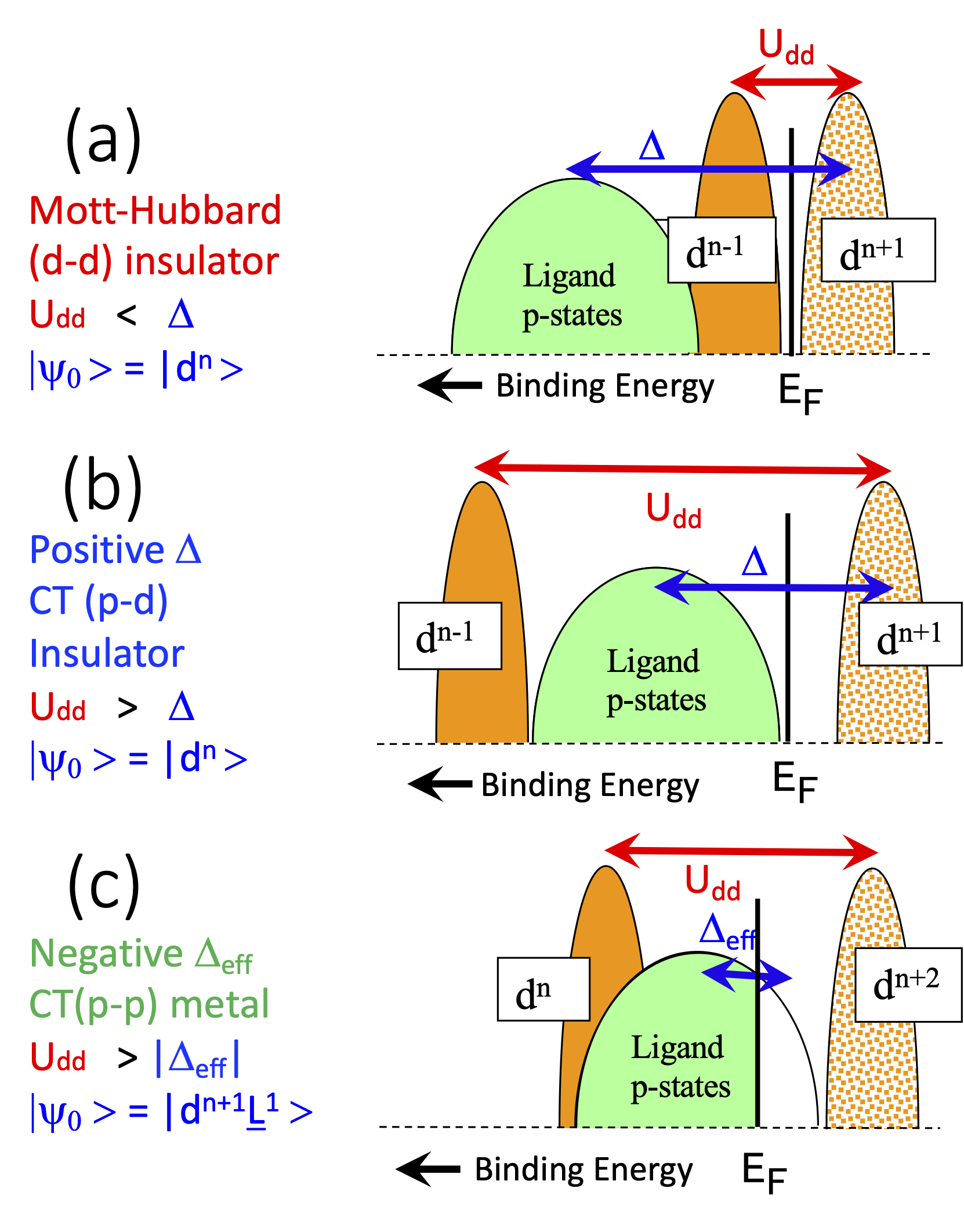}
\caption{ Schematic electronic structure of materials 
representing:
(a) a Mott-Hubbard insulator with $U_{dd}$$<$$\Delta$. A further reduction of $U_{dd}$ would result in a Mott-Hubbard metal if the lower(occupied) and upper(unoccupied) Hubbard $d$-bands overlap. 
(b) a positive-$\Delta$ charge-transfer insulator with $U_{dd}$$>$$\Delta$ and 
\textit{p}$\rightarrow$\textit{d} type lowest energy excitations
(c) an effective negative-$\Delta$ metal with $U_{dd}$$>$$|\Delta_{eff}|$ and $E_F$ positioned within the ligand-$p$ band states, facilitating band inversion with \textit{p}$\rightarrow$\textit{p}-type lowest energy excitations to make it a correlated topological metal.
Note that for Mott-Hubbard and positive-$\Delta$ materials, the ground state $|\psi_0$$>$ has a dominantly $|d^{n}$$>$ character, while for an effective negative-$\Delta$ material, $|\psi_0$$>$ has a dominantly $|d^{n+1}\underline{L}^1$$>$ character, where $\underline{L}^1$ is a hole in the ligand $p$ band.
}
\end{figure}

The present results thus show that while $U_{dd}$ gets reduced by $\sim$40\% compared to CoO, it is still larger than $|\Delta|$. It turns out that the reduced values of $U_{dd}$ = 3.0 eV eV for CoTe$_{2}$ is just right, not too large and not too small. If $U_{dd}$ was too small, CoTe$_{2}$ would have become a $d$-band Mott-Hubbard system in the ZSA picture, as shown schematically in Fig. 12(a). In contrast, if $U_{dd}$ was too large, CoTe$_{2}$ would likely become a positive-$\Delta$ material like CoO (see Fig. 12(b)). But for the moderate $U_{dd}$ as obtained for CoTe$_{2}$ from our experimental results and data analysis, it results in an appropriate value of $U_{dd}$$>$$|\Delta_{eff}|$. This condition allows Te $p$-band states to exhibit an effective negative-$\Delta$ with $E_F$ positioned within the Te-$p$ band states (see Fig. 12(c)). It results in a correlated metal with a narrow Co $3d$ band centered at 0.8 eV BE below $E_F$ (Fig. 7). It simultaneously also facilitates band inversion between Te $p_x$+$p_y$ and Te $p_z$ orbitals just below and above $E_F$\cite{Chakraborty2023}. Thus, the ligand Te 5$p$ states are sandwiched between the occupied and unoccupied $3d$ states in CoTe$_{2}$ and exhibit \textit{p}$\rightarrow$\textit{p}-type lowest energy excitations. This picture fits well with Te $5p$ character bands at and near $E_F$  seen in ARPES of CoTe$_2$.\cite{Chakraborty2023} Further, as discussed in the introduction,  band structure calculations for CoTe$_2$ \cite{Chakraborty2023} showed that the bulk character Type-II Dirac point lies $\sim$0.9 eV above $E_F$. Also, in combination with ARPES studies, it was shown that the Dirac points in surface states of CoTe$_2$ lie $\sim$0.5 eV below $E_F$. Thus, the Dirac points are somewhat away from $E_F$. However, an appropriate modification of the topological states via bandwidth-control or doping on the transition metal-site or ligand-site can lead to tuning a Dirac point closer to $E_F$. This can help to achieve robust topological transport properties. 

\section{\label{sec:levelI}Conclusions}
In conclusion, we could quantify $U_{dd}$, $\Delta$ and $T_{eg}$ in the topological metal CoTe$_2$. The results show a negative-$\Delta$ for CoTe$_2$, compared to a positive-$\Delta$ for the CT insulator CoO. The weaker $T_{eg}$ in CoTe$_2$ compared to CoO rules out $T_{eg}$ as a cause of $U_{dd}$ reduction. The obtained increase in charge-transfer is attributed to negative-$\Delta$ and a reduced $U_{dd}$. However, only because $U_{dd}$$>$$|\Delta|$, CoTe$_{2}$ becomes a topological metal with \textit{p}$\rightarrow$\textit{p} type lowest energy excitations. The study reveals the nexus between negative-$\Delta$ and reduced $U_{dd}$ 
with $E_F$ positioned within the Te-$p$ band states. This results in a correlated metal and simultaneously facilitates band inversion between Te $p_x$+$p_y$ and Te $p_z$ orbitals for achieving topological behavior in CoTe$_2$.

\begin{acknowledgments}
This work was supported by the National Science and Technology Council(NSTC) of Taiwan under Grant Nos. NSTC 113-2112-M-006-009-MY2 (CNK), 110-2124-M-006-006-MY3 (CSL), 112-2124-M-006-009 (CSL), 113-2112-M-007-033 (AF),  112-2112-M-213-029(AC) and 114-2112-M-213-021(AC). AF acknowledges support from the Yushan Fellow Program under the Ministry of Education of Taiwan and Grant No. JP22K03535 from Japan Society for the Promotion of Science(JSPS).  ARS thanks the National Science and Technology Council(NSTC) of Taiwan for a post-doctoral fellowship under Grant No. NSTC 114-2811-M-213-006. 
\end{acknowledgments}

\newpage


 
\setcounter{equation}{0}
\setcounter{figure}{0}
\setcounter{table}{0}
\setcounter{page}{1}
\renewcommand{\theequation}{A\arabic{equation}}
\renewcommand{\thefigure}{S\arabic{figure}}
\renewcommand{\bibnumfmt}[1]{[R#1]}
\renewcommand{\citenumfont}[1]{R#1}


\title{Supplementary Material for 
``The nexus between negative charge-transfer and reduced on-site Coulomb energy in a correlated topological metal CoTe$_2$"}

\author{A. R. Shelke}
\author{C.-W. Chuang}
\author{S. Hamamoto}
\author{M. Oura}
\author{M. Yoshimura}
\author{N. Hiraoka}
\author{C.-N. Kuo}
\author{C.-S. Lue}
\author{A. Fujimori}
\author{A. Chainani}


Supplementary Information contains following Supplementary Notes:

SN1: Optimization of electronic parameters in cluster model calculations

~~~~~1. The $2p$ PES and $L$-edge XAS spectra of CoTe$_2$ 

~~~~~2. The $2p$ PES and $L$-edge XAS spectra of CoO


\maketitle



\subsection{SN1: Optimization of electronic parameters in cluster model calculations}

Charge-transfer cluster model spectral calculations as described in the Methods section were carried out to obtain optimal electronic parameters which can suitably describe
the $2p$ PES and $L$-edge XAS experimental spectra of CoTe$_2$ and CoO, as discussed in the main paper (Fig. 8(a-d)). The full set of optimal parameters are listed in SM Table I for  CoTe$_2$ and CoO. It is clear that the electronic parameters of CoTe$_2$ and CoO are quite different with a negative and positive $\Delta$, respectively. In order to validate the obtained electronic parameters, we carried out a systematic and extensive set of calculations to confirm their optimal values and we discuss a few of them in the following. It is noted that the present study is the first to report a comparison of the $2p$ PES and $L$-edge XAS experimental and calculated spectra to obtain electronic parameters of CoTe$_2$. On the other hand, CoO has been analyzed earlier\cite{Elp2,Ghiasi2} and the parameters for CoO that we obtained are quite close to an analysis using a cluster model\cite{Elp2} as well as a charge-transfer multiplet calculation combined with DMFT method\cite{Ghiasi2}, as shown in SM Table I. 

\subsubsection{1. Simultaneous optimization of the $2p$ PES and $L$-edge XAS spectra of CoTe$_2$}

\begin{figure}
\includegraphics[scale=0.2]{FS1.png}
\caption{(a,b) Simultaneous optimization of charge-transfer energy $\Delta$ for (a) Co 2\textit{p} PES and (b) Co $L$-edge XAS of CoTe$_2$. (c) Attempt at optimization of F\textsubscript{k} from 0.8 to 0.1 by keeping G\textsubscript{k} = 0.8 for Co 2\textit{p} PES.
(d) Attempt at optimization of G\textsubscript{k} from 0.8 to 0.1 by keeping F\textsubscript{k} = 0.8 for Co 2\textit{p} PES.
(e,f) Simultaneous optimization of F\textsubscript{k} and G\textsubscript{k} from 0.8 to 0.1 for (e) Co 2\textit{p} PES
and (f) Co $L$-edge XAS.}
\end{figure}

In Fig. S1(a) and (b), we plot a series of Co $2p$ PES and $L$-edge XAS calculated spectra(blue lines) for CoTe$_2$, respectively, and compare it with experiment(red symbols) for checking the optimal value of 
the charge transfer energy $\Delta$. We varied $\Delta$ = -3.0 to +2.0 eV in 1 eV steps, keeping all other parameters fixed to optimal values.
In particular, the results show that the Co $2p_{3/2}$ PES and $L$-edge XAS satellite feature in the calculated spectra show the least deviation compared to experiment for 
$\Delta$ = -2.0 eV.

Similarly, Fig. S1(c) and (d) shows a series of Co 2p PES calculated spectra(blue lines) for CoTe$_2$ for checking the optimal value of the Slater parameters F$_k$ and G$_k$ independently i.e. by varying only one of them at a time by a reduction factor from 0.8 to 0.1, keeping the other fixed to a reduction factor of 0.8, which is the standard reduction value as discussed in the Methods section. However, it was found that changing only of them, either F$_k$ or G$_k$, did not give a suitable match to experimental data. In particular, the Co $2p_{3/2}$ satellite feature showed significant deviations from experimental spectra(red symbols). It was found that we need to change both of them simultaneously to minimize the deviation of the Co $2p_{3/2}$ satellite feature compared to experiment. The results of such an exercise is shown in Fig. S1(e) and (f), where we
plot a series of  Co $2p$ PES and $L$-edge XAS calculated spectra(blue lines) for CoTe$_2$, respectively, compared with experiment (red symbols). We varied F$_k$ and G$_k$ together from a reduction factor of 0.8 to 0.1, keeping all other parameters fixed to optimal values. The results show that a reduction factor of 0.5 for F$_k$ and G$_k$ shows the least deviation compared to experiment for the Co $2p$ PES and $L$-edge XAS spectra.

\subsubsection{2. Simultaneous optimization of the $2p$ PES and $L$-edge XAS spectra of CoO}

\begin{figure}
\includegraphics[scale=0.18]{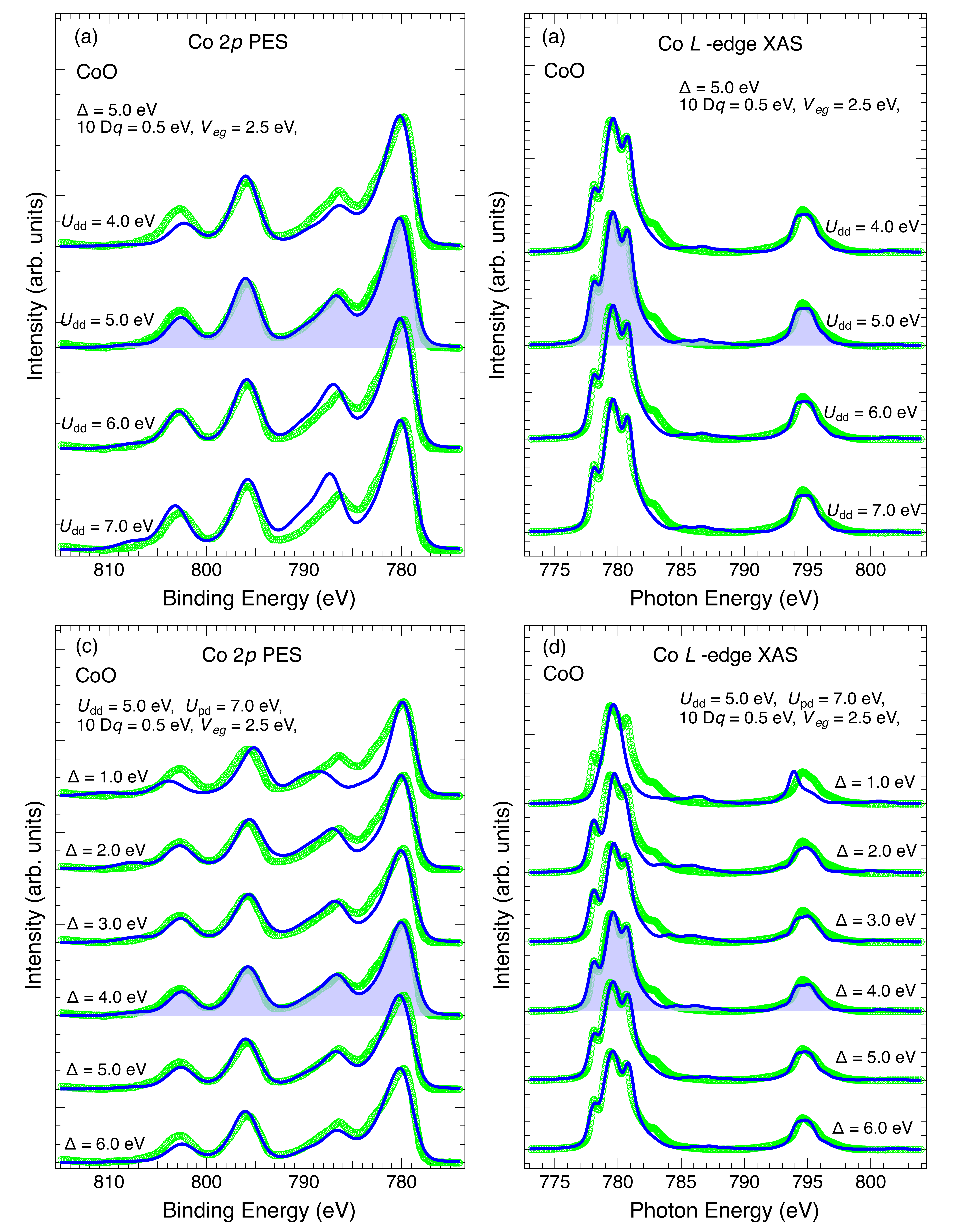}
\caption{(a,b) Simultaneous optimization of on-site energy $U_{dd}$ for Co 2\textit{p} PES and Co $L$-edge XAS of CoO.
(c,d) Simultaneous optimization of charge-transfer energy $\Delta$ for (c) Co 2\textit{p} PES and (d) Co $L$-edge XAS of CoO.}
\end{figure}

Fig. S2(a) and (b) shows a series of Co $2p$ PES and $L$-edge XAS calculated spectra(blue lines) for CoO, respectively, for checking the optimal value of the 
the on-site Coulomb energy $U_{dd}$ and compare it with experiment(green symbols). We varied $U_{dd}$ = 4.0 to 7.0 eV in 1 eV steps, keeping all other parameters fixed to optimal values. The experimental Co $2p$ HAXPES spectrum is taken from ref\cite{AC2}.
The results show that, compared to the optimal value of  $U_{dd}$ = 5.0 eV, the Co $2p_{3/2}$ satellite feature in the calculated spectra deviates from experiment for larger $U_{dd}$ values. On the other hand, both, the Co $2p_{3/2}$ and Co $2p_{1/2}$ satellite features in the calculated spectra together deviate from experiment for smaller $U_{dd}$ values compared to the spectrum obtained with the optimal value of  $U_{dd}$ = 5.0 eV.

Fig. S2(c) and (d) shows  Co $2p$ PES and $L$-edge XAS calculated spectra(blue lines) for CoO, respectively, for checking the optimal value of the 
the charge transfer energy $\Delta$ and compare it with experiment(green symbols). We varied $\Delta$ = 1.0 to +6.0 eV in 1 eV steps, keeping all other parameters fixed to optimal values.
It is observed in Fig. S2(c) that the Co $2p_{3/2}$ satellite feature in the calculated spectra shows least deviation for $\Delta$ = 4.0 eV. But shows deviations from experimental main peak widths and satellite widths and intensities for  $\Delta$$<$4.0 eV, and to a lesser extent also for $\Delta$$>$4.0 eV. A similar behavior for  $\Delta$$<$4.0 eV is also seen in the Co $L$-edge XAS spectra for CoO shown in Fig. S2(d). However, the Co $L$-edge XAS spectrum for $\Delta$ = 4.0 eV shows the least deviation in the leading edge prepeak/multiplet compared to also the spectra obtained for $\Delta$$>$4.0 eV. Taken together, the simultaneous optimization of Co $2p$ PES and $L$-edge XAS spectra with $\Delta$ = 4.0 eV is considered to show least deviation compared to experiment.

\begin{table}[t!]
	
	 \caption{Electronic parameters and $d^n$ count for CoTe$_2$ and CoO using  3-basis state cluster model calculations.}
\begin{tabular}{ccccc}

\hline
~~~~& ~~~CoTe$_2$~~~&~~~CoO~~~&~~~CoO~~~&~~~CoO~~~\\
~~~~&~~~~&~~~~&~ref.\cite{Elp}~&~ref.\cite{Ghiasi}~\\
Parameter        &~~~~&~~~~\\
\hline
$U_{dd}$(eV)		&~3.0~&~5.0~&~5.3~&~6.8~ \\
$\Delta$(eV)		&~-2.0~&~4.0~&~5.5~&~4.1~\\

$T_{eg}$(eV)		&~1.2~&~2.5~&~2.25~&~2.0~\\
$T_{t2g}$(eV)		&~0.6~&~1.25~&~1.0~&~1.2~\\

$10Dq$(eV)		&~1.0~&~0.5~&~0.7~&~0.25~\\

$F_k$,$G_k$		&~0.5~&~0.8~&~0.8~&~0.8~\\

$U_{dd}/T_{eg}$	&~2.5~&~2.0~&~2.3~&~3.4~\\

$\big|\Delta\big|/T_{eg}$&~1.7~&~1.6~&~2.4~&~2.05~\\

$d^n$ count		&~8.14~&~7.21~&~7.22~&~--~\\
\hline
\end{tabular}
	
\end{table}

\end{document}